\documentclass[twocolumn]{aastex63}

\usepackage{multirow}
\usepackage{lineno}
\usepackage{amsmath}
\usepackage{comment}
\usepackage[flushleft]{threeparttable}

\newcommand{\ghostH}{$73.22\pm2.06$ km/s/Mpc} 
\newcommand{\ghostcal}{$25.361 \pm 0.0136$} 

\newcommand{\simpleavecal}{$25.40 \pm 0.047$} 

\newcommand{\numsimpleavecal}{$25.47$} 

\newcommand{\ghostsn}{$15.213 \pm 0.049$} 

\newcommand{\medianH}{$72.94\pm1.98$ km/s/Mpc \,} 
\newcommand{\mediancal}{$25.353 \pm 0.014$} 
\newcommand{\mediansn}{$15.213 \pm 0.046$} 

\newcommand{\lowchiH}{$75.24\pm2.21$ km/s/Mpc} 
\newcommand{\lowchical}{$25.374 \pm 0.015$} 
\newcommand{\lowchisn}{$15.164 \pm 0.052$} 

\newcommand{\lowdispH}{$73.39\pm1.83$ km/s/Mpc } 
\newcommand{\lowdispcal}{$25.355 \pm 0.015$} 
\newcommand{\lowdispsn}{$15.201 \pm 0.040$} 

\newcommand{\lowdisp}{$0.18$} 
\newcommand{\lowchi}{$1.00$} 

\newcommand{\sysH}{$0.83\,$} 

\newcommand{\Nminusone}{N_{\mathrm{+,1.0}}}

\definecolor{newpurple}{rgb}{0.7,0,1}

\newcommand{\scanswered}[1]{{}} 

\shorttitle{$H_0$ from Standardized TRGB}
\shortauthors{CATS team}

\begin{document}

\title{CATS: The Hubble Constant from Standardized TRGB and Type Ia Supernova Measurements}

\correspondingauthor{Dan Scolnic}
\email{daniel.scolnic@duke.edu}

\author[0000-0002-4934-5849]{D. Scolnic}
    \affil{Department of Physics, Duke University, Durham, NC 27708, USA}
    \author[0000-0002-6124-1196]{A. G. Riess}
    \affil{Space Telescope Science Institute, Baltimore, MD, 21218, USA}
    \affil{Department of Physics and Astronomy, Johns Hopkins University, Baltimore, MD 21218, USA}
    
    \author[0000-0003-3829-967X]{J. Wu}
    \affil{Kuang Yaming Honors School, Nanjing University, Nanjing, Jiangsu 210023, China}
    \affil{Department of Physics, Duke University, Durham, NC 27708, USA}

\author[0000-0002-8623-1082]{S. Li}
    \affil{Department of Physics and Astronomy, Johns Hopkins University, Baltimore, MD 21218, USA}

 \author[0000-0002-5259-2314]{G. S. Anand}
    \affil{Space Telescope Science Institute, Baltimore, MD, 21218, USA}
\author[0000-0002-1691-8217]{R. Beaton}
    \affil{Space Telescope Science Institute, Baltimore, MD, 21218, USA}
    \affil{Department of Physics and Astronomy, Johns Hopkins University, Baltimore, MD 21218, USA}
    \affil{Department of Astrophysical Sciences, Princeton University, Princeton, NJ 08544, USA}  
\author{S. Casertano}
    \affil{Space Telescope Science Institute, Baltimore, MD, 21218, USA}
    \author[0000-0001-8089-4419]{R.~I. Anderson}
    \affil{Institute of Physics, \'Ecole Polytechnique F\'ed\'erale de Lausanne
(EPFL), 1290 Versoix, Switzerland}
    \author[0000-0002-2376-6979]{S. Dhawan}
    \affil{Institute of Astronomy and Kavli Institute for Cosmology, University of Cambridge, Cambridge CB3 0HA, UK}
\author{X. Ke}
    \affil{Department of Physics, Duke University, Durham, NC 27708, USA}

\begin{abstract}
The Tip of the Red Giant Branch (TRGB) provides a luminous standard candle for constructing distance ladders to measure the Hubble constant. In practice its measurements via edge-detection response (EDR) are complicated by the apparent fuzziness of the tip and the multi-peak landscape of the EDR.  As a result, it can be difficult to replicate due to a case-by-case measurement process.  Previously we optimized an unsupervised algorithm, Comparative Analysis of TRGBs (CATs), to minimize the variance among multiple halo fields per host without reliance on individualized choices, achieving state-of-the-art $\sim$ $<$ 0.05 mag distance measures for optimal data.   Further, we found an empirical correlation at 5$\sigma$ confidence in the GHOSTS halo survey between our measurements of the tip and their contrast ratios (ratio of stars 0.5 mag just below and above the tip), useful for standardizing the apparent tips at different host locations.  Here, we apply this algorithm to an expanded sample of SN Ia hosts to standardize these to multiple fields in the geometric anchor, NGC 4258. In concert with the Pantheon$+$ SN Ia sample, this analysis produces a (baseline) result of $H_0=~$\ghostH. The largest difference in $H_0$ between this and similar studies employing the TRGB derives from corrections for SN survey differences and local flows used in most recent SN Ia compilations but which were absent in earlier studies.  SN-related differences total $\sim$ 2.0 km/s/Mpc.   A smaller share, $\sim$ 1.4 km/s/Mpc, results from the inhomogeneity of the TRGB calibration across the distance ladder.  We employ a grid of 108 variants around the optimal TRGB algorithm and find the median of variants is \medianH with an additional uncertainty due to algorithm choices of \sysH km/s/Mpc.  None of these TRGB variants result in $H_0$ less than 71.6 km/s/Mpc.  
\end{abstract}

\keywords{galaxies: distances and redshifts; cosmology: distance scale}

\section{Introduction}

The Tip of the Red Giant Branch (TRGB) is a waypoint in the evolutionary state for giant stars and offers an emerging tool for measuring extragalactic distances \citep{Lee93,Serenelli:2017,2018SSRv..214..113B}. Distances obtained via the TRGB have proven vital for grounding other physical measurements of nearby galaxies \citep{2021ApJ...918...23M,2021ApJ...914L..12S,2022ApJ...926...77M}, and have led to a more thorough understanding of the structure of our local universe \citep{2022ApJ...927..168S,2023ApJ...944...94T}.  Recently, it has also been employed as a center-piece of some distance ladders used to measure the Hubble constant ($H_0$) \citep{Jang17,Freedman20,Blakeslee21,Anand22,Dhawan22,Jones22}.  Results from this measurement are particularly interesting in light of the intriguing `Hubble tension', which is an empirical difference between direct measures of $H_0$ from the distance-redshift relation measured in the late universe and its predicted value based on the calibration of $\Lambda$CDM in the early, pre-recombination Universe primarily via a calibration of the sound horizon (e.g. \citealp{Planck18}).  Distance ladders which include the TRGB find values between two of the main tentpoles (directly using a ladder that includes Cepheids of $73\pm1$ km/s/Mpc -  \citealp{Riess22}, hereafter R22, and inferred using the Cosmic Microwave Background of $67.4\pm0.5$ km/s/Mpc - \citealp{Planck18}). Some TRGB results (e.g., $\sim70$ km/s/Mpc, \citealp{Freedman20}) are consistent with both to 1.5~$\sigma$ or less, while others are somewhat higher at  $\sim71.5-73$   \citep{Anand22,Blakeslee21,Jones22}.  It is unclear whether these differences are due to statistical fluctuations, differences in TRGB measurements, or arise from other parts of the distance ladder unrelated to the TRGB, but reconciling any potential differences is paramount for improving our understanding of the Hubble tension. 

Our ability to grasp the significance of any differences in TRGB analyses is further complicated by the complexity of the measurement itself.  
The empirical tip is due to the abrupt end of the luminosity function of RGB stars, but in practice, the tip often appears fuzzy due to unavoidable contamination by AGB stars, other younger populations as well as photometric errors and, in some cases, small-number statistics. Edge-detection is equivalent to evaluating an empirical derivative of a noisy function and so it often produces multiple peaks and the potential for ambiguity in the identification of the TRGB. 
Different TRGB measurement methods address this ambiguity using different choices in how the measurement is made.
It is hard to evaluate how individual choices for the identification affect the determination of the Hubble constant without a systematic, algorithmic approach that considers how all such variations in the procedure propagate to $H_0$.  This paper only deals with edge detection such as with a Sobel-like filter as a method to measure the TRGB, and does not touch upon potential systematics in the luminosity-function fitting method \citep{Mendez_2002,Makorov06,Wu14}, which are likely different and outside the scope of this work.

When TRGB measurements are used as part of the distance ladder, they are used in the first two rungs, first for an absolute calibration using a geometric measurement, and second, to calibrate the luminosity of secondary distance indicator like Type Ia supernovae (SN Ia; \citealp{Scolnic22}), Surface Brightness Fluctuations (SBF, \citealp{Blakeslee21}) or the Tully-Fisher relation \citep{Kourkchi22}.  A great deal of recent work has focused on the first rung and the determination of an absolute $I$ magnitude of the TRGB with most results falling within a range of $\sim -4.06<I<-3.95$ mag \citep{Freedman20,Hoyt_2021,Li22,Blakeslee21,Capozzi20}.  

Some differences in the calibration rung are likely astrophysical in nature as differences at the $\sim$0.1-0.2 mag level appear within a single host and appear to depend on the stellar populations being probed in a specific study \citep{Wu22,Anderson23,Hoyt23}; other calibration differences may be due to differences in the measurement process itself.  It is therefore essential to consider how to rectify such differences between the first and second rung rather than selecting the sharpest EDR.    For example, recent papers (e.g. \citealp{Jang21,Hoyt23}) discussed using the optimal fields among many in the LMC or NGC 4258 that are available and then use these optimal fields for calibration. While these choices may be well-motivated to derive the most precise value for calibration purposes, it may create a difference in how the first rung and second rung galaxies are treated within the distance ladder.  More specifically, the criteria used to select ideal fields may not be feasible to apply for more distant second-rung galaxies or the very nature of the difference in photometry quality may produce systematics.

As a first step, in \cite{Wu22} (hereafter W22), we ``trained'' an unsupervised algorithm for measuring the EDR to yield internally accurate and precise TRGB measurements for a wide range of halo fields around a common host.  This is a powerful approach for avoiding confirmation bias of SNe or $H_0$ as the algorithm was optimized independent of a subsequent application to determine the Hubble constant.
In order to optimize the algorithm, we systematically varied the most widely employed algorithmic features for measuring the TRGB, such as spatial filtering to remove star-forming regions, parameters controlling the smoothing and weighting of the luminosity function, the use of color selection to define the RGB region, and the selection of the maximum in the response function identifying the tip.  For the last element, we found that a very informative quantity is the contrast ratio $ R $ at the tip, i.e., the ratio of the number of stars 0.5 magnitudes below vs. above the tip.  We varied the algorithm parameters to measure the TRGB in multiple halo fields for each of the galaxies observed as part of the GHOSTS program \citep{Radburn_Smith_2011}; the goal was to identify the selections that optimize the consistency of the tip measurements across different fields for the same galaxy, without prejudice regarding its actual brightness.

By this process, we recognized an empirical relation with 5$\sigma$ confidence between the observed tip magnitude of a field and the measured contrast ratio $ R$, defined above, that would produce apparent differences in the tip at the $0.05-0.1$ mag level for typical fields.  Differences in tip magnitudes of this order (at 5$\sigma$ significance) have been reported based on variability-selected subsamples in the LMC by \cite{Anderson23} and could point to population effects, e.g., related to age or metallicity.  Stellar population synthesis models indicate that $R$ is a function of both age and metallicity (W22) with younger populations containing more AGB stars (brighter than the tip) and lower values of $R$.  We found that we are able to reach a dispersion as low as $0.03$ mag per field (1.5\% in distance) for very high $R$, and the dispersion increases rapidly as $R$ decreases, showing that $R$ is both an indicator of the quality of a tip measurement and its brightness.  A standardization technique for the application of the TRGB may improve the consistency between the first and second rungs. This is important for eliminating the difficulty and vagaries of choosing which fields between rung one and two are comparable.

We use the maser host NGC 4258 as our sole source of TRGB geometric calibration because its TRGB can be measured most consistently with that in SN Ia hosts.  The LMC has too large of an angular extent to measure efficiently with the Hubble Space Telescope (HST), which is essential for a self-consistent TRGB calibration.  A calibration based on the Milky Way is challenging because MW stars are not at a single uniform distance, although several attempts have produced useful results \citep{Li22,Soltis21,Freedman21}. We utilize HST photometry of host galaxies provided by the Color Magnitude Diagrams/Tip of the Red Giant Branch Catalog (CMDs/TRGB) within the Extragalactic Distance Database (EDD; \citealp{Tully2009,Anand21b}), which provides photometry with uniform procedures across the full range of hosts and calibrators, effectively reducing sensitivity of $H_0$ to experimental differences in photometry or flux-scale calibration.  The companion paper by Li et al. 2023 (in prep., hereafter L23) presents an analysis focused on the measurements in NGC 4258.  

The structure of this paper is as follows.  In section II, we present the color-magnitude diagrams and other properties of the TRGB fields in supernova host galaxies, and derive the tip brightnesses of each field.  In section III, we describe the methodology to determine $H_0$ and present our baseline results.  In section IV, we discuss variants in our analysis method and their impact on inferences of $H_0$.  In section V we discuss comparison to literature results and in section VI we present our conclusions.

\section{Measurements}

\begin{figure*}
    \centering
    \includegraphics[width=0.9\textwidth]{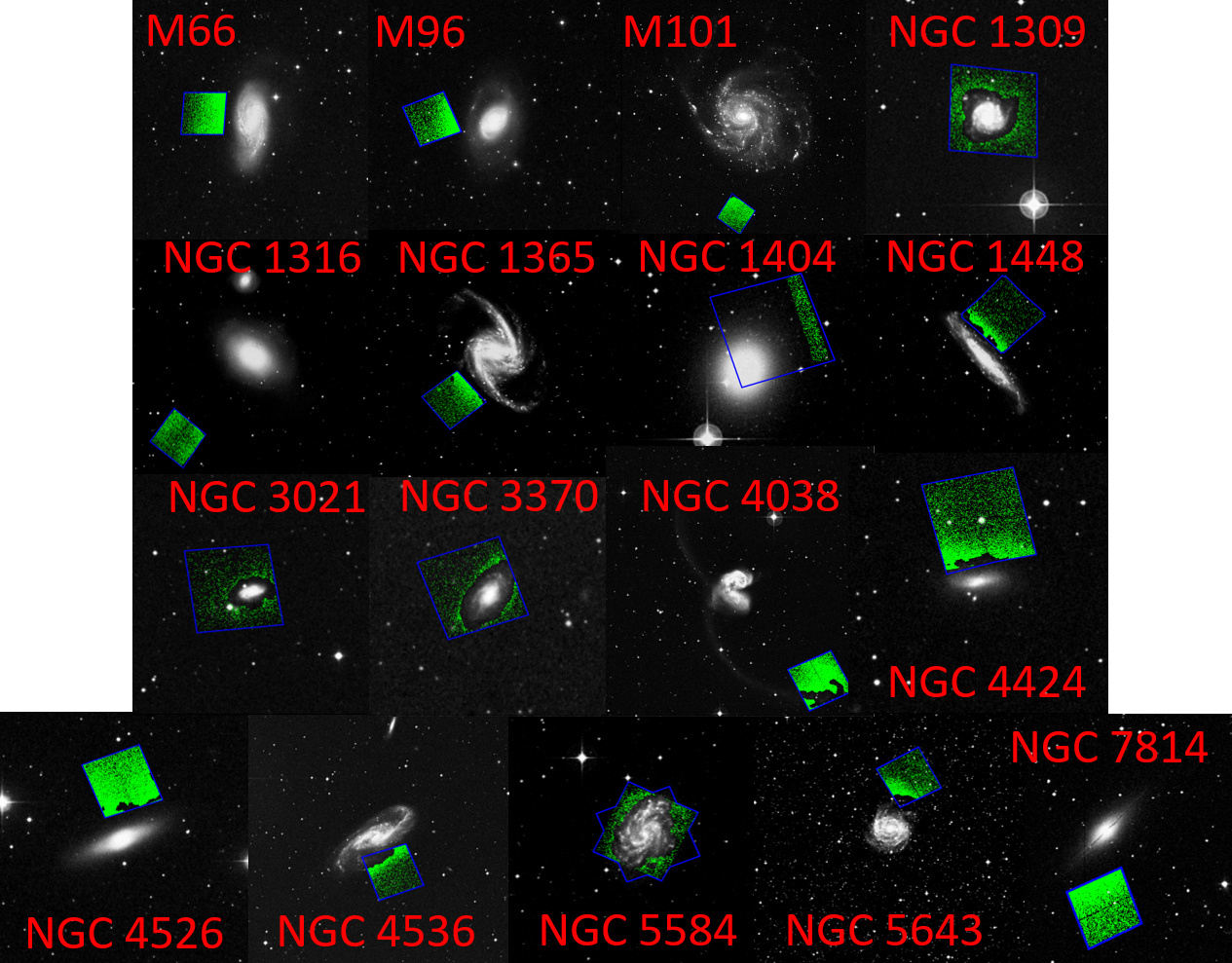}
    \caption{Footprints  (shown with a rectangular outline) of the observed fields around galaxies used in the second rung of the distance ladder. The background images are from SDSS \citep{SDSS20}. A green dot is drawn for each star that is included in the analysis of the CMD with blank regions due to spatial masking of regions with high blue (young) star density as discussed in Section 2.1. An exception is NGC 1404 where we followed \cite{Hoyt_2021} who showed the field is severely crowded and we only consider a similar, halo strip.}
    \label{fig:fields}
\end{figure*}

\begin{figure*}
    \centering
    \includegraphics[width=0.9\textwidth]{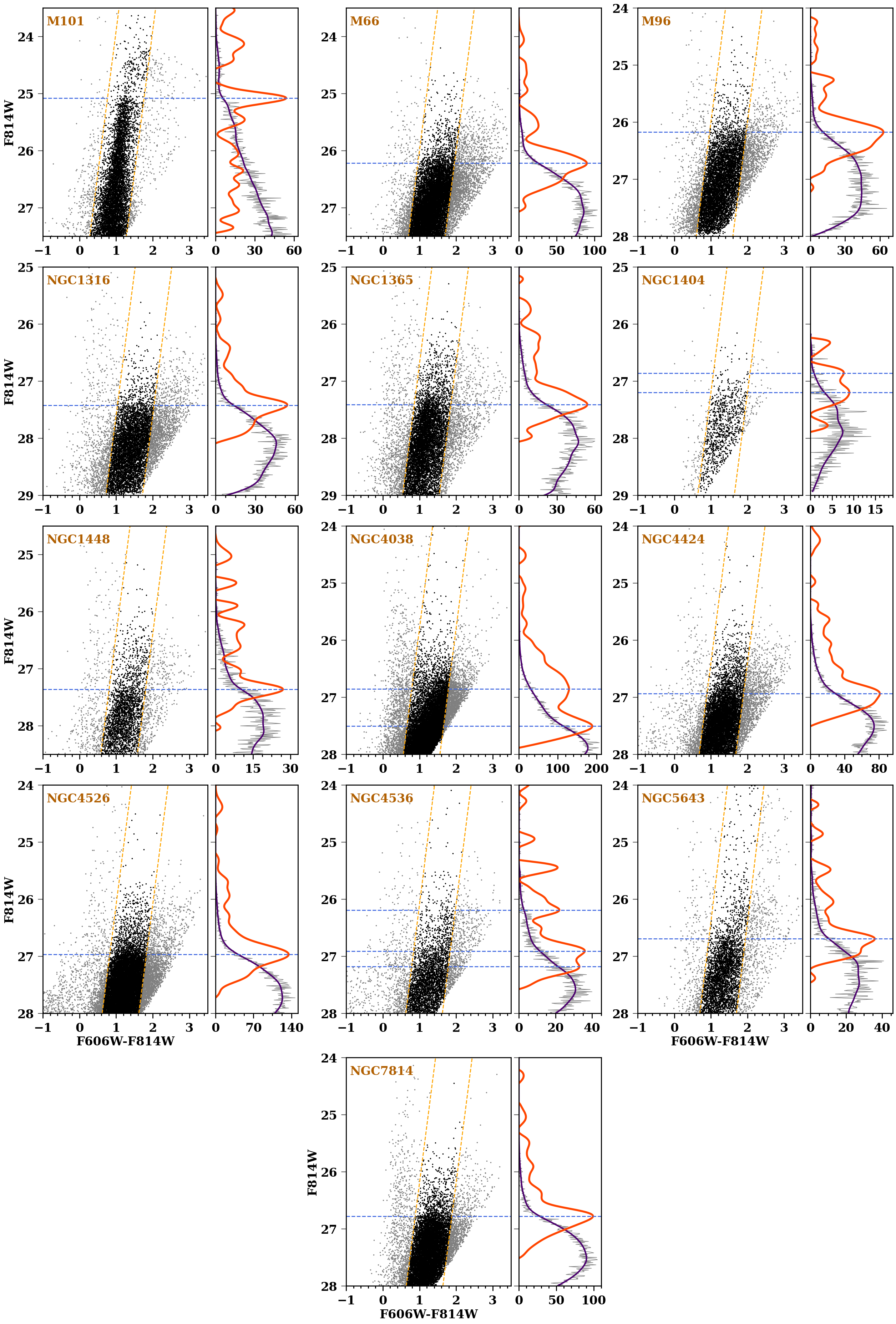}
    \caption{The color-magnitude diagrams (CMDs) of SN fields with significant tip detections ($R>2$). On the left panel of each subplot, the orange lines show the color bands (i.e., selection region) for each field with the baseline 1.0 mag width, and the blue line(s) shows the TRGB detection(s). On the right panel, the purple curve shows the smoothed luminosity function with baseline smoothing scale of $0.1$ mag smoothing width, while the red curve shows the edge-detection response.}
    \label{fig:CMD}
\end{figure*}

\subsection{CATs in SN Ia hosts}

To calibrate the SN Ia fiducial luminosity with the TRGB, we compiled a list of all SN Ia hosts for which deep halo imaging exists with HST ACS in $F814W$ and a bluer color ($F606W$ or $F555W$), with the host imaging and SN measurements available at the time of this analysis. The majority of the hosts have already been included in the studies by both \cite{Freedman19b} (CCHP) and \cite{Anand21} (EDD; see Table 1 of both papers respectively).  There are two SNe that we add from the year 2021; a second SN in NGC 1448 (SN 2021pit, R22) and SN 2021rhu in NGC 7814 \cite{Dhawan22}. Thus the sample studied in previous papers includes a total of 17 galaxies host to a total of 22 SNe, providing the second rung of the distance ladder. Most of the data were obtained explicitly to measure the TRGB in the halo of the respective galaxies, although four hosts from the SH0ES programs (R22) were observed with the primary goal of obtaining Cepheid observations, and had TRGB measurements performed by \cite{Jang_2018}. Two additional SN host galaxies, NGC 4038 and NGC 7814, were observed by HST programs 10889 and 10580 (I. Savianne). A parallel analysis of HST ACS images of NGC 4258 is carried out in L23.

As can be seen in Fig.~\ref{fig:fields}, the locations of the pointings relative to the disk/halo vary greatly among the different galaxies; while some pointings are well into the halo (e.g. M101, NGC 1316), others have moderate to complete overlap with the disks or bulge (e.g. NGC 3021, NGC 1309, NGC 3370, NGC 5584, NGC 1404, NGC 5643). For the purpose of limiting the TRGB measurement to exclusively older populations, these images are not all strictly from the halo, and hence demonstrate the necessity of using an algorithm for excluding younger regions which we can associate with blue, main sequence stars.  We only analyze the photometry from \cite{Anand21}, as the photometry used by \cite{Freedman19b} has not been made publicly available.  \cite{Anand21} performs PSF photometry with DOLPHOT \citep{DOLPHOT}.  \cite{Jang22} studies differences in photometry of the TRGB when using the DOLPHOT package or the DAOPHOT package (\citealp{Stetson87}, the method used by CCHP) and shows there can be agreement better than 0.02 mag, though the exact level of agreement depends on the specific reduction parameters used, as well as corrections applied in the post-processing (e.g. aperture corrections). Thus, it is important to use a homogeneous photometry reduction within a single study.

Color-magnitude diagrams for all the galaxies for which we can assess the tip are shown in Fig.~\ref{fig:CMD}.  To calculate the edge detection response (seen on on the right of each panel from Fig.~\ref{fig:CMD}), we employ the same automated algorithm to measure the TRGB here that was developed and optimized in W22 to produce the lowest TRGB dispersion for multiple fields in the same host for the GHOSTS sample.  As seen in the CMDs of Fig.~\ref{fig:CMD}, the sharpness of the TRGB peak varies significantly across fields, so it is important to maintain a robust algorithm for identifying the tip independent of external knowledge of the distance.  We identify four primary features of a TRGB measurement algorithm, each of which is associated with a parameter that may be globally optimized.  We describe them step-wise below; for all steps we adopt the procedures discussed in W22.

\subsubsection{Spatial Selection}

Our first step is to perform a spatial selection of a usable region of a frame by automated detection and exclusion (masking) of young/blue regions, parametrized by a frame-specific threshold.  We identify the peak density of blue stars and then exclude regions of the frame whose density exceeds a modest fraction of this value, i.e., greater than $\textbf{\textrm{SpatialClip\%}}$, where the blue star density is calculated in units of number of stars per 16 square arcseconds. From W22, $\textit{\textrm{SpatialClip\%}}$ is set to be 10\% of the frame maximum.  A global minimum of 4 for the peak is used to contend with cases that have negligible blue stars because they are already far out in the halo.  We identify the blue stars as those bluer than 0.3 mag in F606W-F814W color, or 0.6 mag in F555W-F814W color. We adopt a significance threshold for detections of SNR $> 4$, lower than the value $ > 10 $ used in W22; the reason is that this sample is generally at greater distances than the GHOSTS sample studied in W22, and that photometric uncertainties are evaluated differently by \cite{Anand22} than in the GHOSTS analysis \citep{Radburn_Smith_2011}.

The impact of the clipping is shown in Fig.~\ref{fig:fields}, and has the intended effect of reducing the overlap of the pointings with the disk area of each galaxy.  The range of fractional field clipping is $\sim5\%-20\%$ for SN fields and $\sim20\%-40\%$ for NGC 4258 fields. 
In \S 4 we also present results obtained with double or half the clipping threshold.
The typical number of stars is $\sim$ 10000 for SN fields and $\sim$ 5000 for NGC 4258 fields, due to the different distances.

\subsubsection{CMD Range Selection}

Second, we select a diagonal band of the CMD encompassing the RGB to measure the LF. We introduce a parameter representing the width of the color band as  $\textbf{\textrm{Width}}$. The optimal value of $\textit{Width}$ for the GHOSTS sample was found by W22 to be 1.0~mag.  The center and slope of the band is varied for each frame to maximize the number of stars in the band. For most fields (and all those used to measure $H_0$) both F606W and F814W data are available. We discuss four galaxies for which only F555W and F814W are available (bottom row of Fig.~\ref{fig:CMD} in more detail in the Appendix. The range of allowed slopes is $ -7 $ to $ -5 $ for the CMDs with $F814W$ versus $F606W-F814W$, and different slopes can be seen for the color bands in Fig.~\ref{fig:CMD}.

\subsubsection{Smoothing the Luminosity Function}
The next step is to build a smooth luminosity function.  We introduce a common parameter $\textbf{\textrm{Smooth}}~\sigma$ which is size of the kernel for Gaussian smoothing. After creating the luminosity function from the CMD, we smooth the luminosity with a GLOESS with smooth $\sigma$ mag for all galaxies. From W22, the optimal $\textit{\textrm{Smooth}}~\sigma$ is set to be 0.1 mag.

\subsubsection{Measuring the EDR}
Finally, we measure the peaks of the EDR.  We introduce a parameter  $\textbf{\textrm{MinTh~\%}}$ which is the threshold relative to the peak edge detection for which other peaks would be included in the analysis.  All tips are measured using a Sobel filter and Poisson weighting as described in \cite{Hatt_2017}. We retain all measured tip brightnesses whose EDR is greater than $\textit{\textrm{MinTh}~\%}$ of the brightest peak detected.  From W22, $\textit{\textrm{MinTh~\%}}$ is set to 60\%.  For each tip in each field, we measure its magnitude and contrast ratio, $R$, defined as the ratio of the number of selected stars that are 0.5 mag below the tip (i.e., in the RGB region) to those 0.5 mag above the tip (i.e., in the AGB region). In the next section we will introduce the possibility of rejecting tips which are formal outliers among a universal relation (e.g., supernova vs TRGB) which can contend with an unrelated feature like the tip of the AGB. We also define a value $N_{\mathrm{+,1.0}}$ as the number of stars 1 mag below the tip.

Additionally, we correct the magnitude of each field by its MW extinction given by \cite{Schlafly11}. For further tests, we also measure the mean color of the peak magnitude, the number of stars in the field, and the depth.  We provide all these values in Table~\ref{tab:TRGBresults}. Furthermore, following W22, we also calculated the expected internal extinction based on (\citealp{Menard10}; Eq. 30).  Due to the typical distances of fields from the center of the disk, the internal extinction estimates are generally $\leq$ 0.02 mag and have negligible impact on our conclusions (effectively cancelling between the SN hosts and the fields of NGC 4258, see also \citealp{Anderson2022}).

\begin{deluxetable*}{ |c c c c c | c c c | c c c |} 
\tabletypesize{\scriptsize}
\tablecaption{Summary of TRGB parameters and field characteristics for baseline case}
\label{tab:TRGBresults}
\tablehead{\colhead{Galaxy} & \colhead{Field} &  \colhead{Field-RA} & \colhead{Field-DEC} & \colhead{Dist.} & \colhead{$m_{I,TRGB}$}  & \colhead{$R$} & \colhead{$N_{\mathrm{+,1.0}}$} & \colhead{Tip Color} & \colhead{MW Ext.} & \colhead{Int. Ext.}} 
\startdata
SN Fields: & ~ & ~ & ~ & ~ & ~ & ~ & ~ & ~ & ~ & ~\\
M101 & M101 & 210.828 & 54.156 & 22.54 & $25.080\pm0.111$ & 4.4 & 1374.0 & 1.30 & 0.016 & 0.016\\
M66 & M66 & 170.131 & 12.996 & 24.02 & $26.219\pm0.041$ & 6.1 & 7000.0 & 1.54 & 0.060 & 0.015\\
M96 & M96 & 161.768 & 11.825 & 19.11 & $26.171\pm0.040$ & 6.7 & 3420.0 & 1.53 & 0.046 & 0.018\\
NGC1316 & NGC1316 & 50.802 & -37.323 & 70.92 & $27.421\pm0.041$ & 6.1 & 3682.0 & 1.56 & 0.038 & 0.006\\
NGC1365 & NGC1365 & 53.464 & -36.201 & 35.76 & $27.411\pm0.204$ & 3.8 & 4079.0 & 1.33 & 0.037 & 0.011\\
NGC1404 & NGC1404 & 54.695 & -35.568 & 10.31 & $26.860\pm0.040$ & 7.4 & 451.0 & 1.47 & 0.020 & 0.031\\
NGC1404 & NGC1404 & 54.695 & -35.568 & 10.31 & $27.200\pm0.152$ & 4.2 & 624.0 & 1.56 & 0.020 & 0.031\\
NGC1448 & NGC1448 & 56.111 & -44.604 & 63.17 & $27.362\pm0.419$ & 3.2 & 1768.0 & 1.33 & 0.025 & 0.007\\
NGC4038 & NGC4038 & 180.360 & -18.989 & 81.10 & $26.857\pm0.192$ & 3.8 & 9564.0 & 1.44 & 0.084 & 0.005\\
NGC4038 & NGC4038 & 180.360 & -18.989 & 81.10 & $27.507\pm0.580$ & 2.3 & 12164.0 & 1.26 & 0.084 & 0.005\\
NGC4424 & NGC4424 & 186.802 & 9.459 & 25.52 & $26.938\pm0.151$ & 4.0 & 6249.0 & 1.43 & 0.038 & 0.014\\
NGC4526 & NGC4526 & 188.516 & 7.753 & 33.82 & $26.969\pm0.043$ & 5.6 & 10368.0 & 1.37 & 0.040 & 0.011\\
NGC4536 & NGC4536 & 188.588 & 2.145 & 25.58 & $26.193\pm0.200$ & 4.0 & 868.0 & 1.57 & 0.033 & 0.014\\
NGC4536 & NGC4536 & 188.588 & 2.145 & 25.58 & $26.913\pm0.364$ & 3.3 & 2517.0 & 1.40 & 0.033 & 0.014\\
NGC4536 & NGC4536 & 188.588 & 2.145 & 25.58 & $27.183\pm0.676$ & 2.3 & 2531.0 & 1.33 & 0.033 & 0.014\\
NGC5643 & NGC5643 & 218.142 & -44.115 & 19.84 & $26.694\pm0.178$ & 4.0 & 2345.0 & 1.46 & 0.306 & 0.018\\
NGC7814 & NGC7814 & 0.814 & 16.071 & 28.83 & $26.780\pm0.041$ & 6.2 & 7498.0 & 1.40 & 0.081 & 0.013\\
\hline
NGC 4258 Fields: & ~ & ~ & ~ & ~ & ~ & ~ & ~ & ~ & ~ & ~\\

NGC4258 & NGC4258-1 & 184.829 & 47.333 & 18.49 & $25.360\pm0.040$ & 7.0 & 3193.0 & 2.07 & 0.030 & 0.019\\
NGC4258 & NGC4258-1 & 184.829 & 47.333 & 18.49 & $25.690\pm0.462$ & 2.9 & 4529.0 & 2.00 & 0.030 & 0.019\\
NGC4258 & NGC4258-10 & 184.607 & 47.209 & 33.16 & $25.268\pm0.040$ & 11.4 & 531.0 & 1.33 & 0.030 & 0.011\\
NGC4258 & NGC4258-2 & 184.673 & 47.492 & 26.27 & $25.367\pm0.108$ & 4.4 & 1971.0 & 2.25 & 0.030 & 0.014\\
NGC4258 & NGC4258-3 & 184.840 & 47.450 & 34.81 & $25.334\pm0.130$ & 4.4 & 480.0 & 1.98 & 0.030 & 0.011\\
NGC4258 & NGC4258-4\_G1 & 184.879 & 47.353 & 29.16 & $25.304\pm0.048$ & 5.4 & 1386.0 & 2.04 & 0.030 & 0.013\\
NGC4258 & NGC4258-4\_G2 & 184.918 & 47.320 & 33.05 & $25.288\pm0.041$ & 6.4 & 902.0 & 2.06 & 0.030 & 0.012\\
NGC4258 & NGC4258-5 & 184.641 & 47.249 & 22.93 & $25.441\pm0.040$ & 6.7 & 2451.0 & 1.42 & 0.030 & 0.016\\
NGC4258 & NGC4258-5 & 184.641 & 47.249 & 22.93 & $25.781\pm0.567$ & 2.6 & 3199.0 & 1.32 & 0.030 & 0.016\\
NGC4258 & NGC4258-6 & 184.901 & 47.243 & 25.93 & $25.462\pm0.040$ & 6.4 & 4120.0 & 1.41 & 0.030 & 0.014\\
NGC4258 & NGC4258-7 & 184.852 & 47.323 & 21.56 & $25.280\pm0.040$ & 7.5 & 3550.0 & 1.48 & 0.030 & 0.016\\
NGC4258 & NGC4258-8 & 184.978 & 47.237 & 38.87 & $25.302\pm0.040$ & 7.2 & 474.0 & 1.29 & 0.030 & 0.010\\
NGC4258 & NGC4258-9 & 184.977 & 47.332 & 44.33 & $25.320\pm0.041$ & 6.3 & 302.0 & 1.34 & 0.030 & 0.009\\

\enddata
\tablecomments{The Dist. is given to the center of the galaxy in kpc.  TRGB magnitudes are raw and are not corrected for extinction in this column. The $m_{I,TRGB}$ values are the measured tip brightnesses, not the standardized values. $R$ is the contrast ratio and $N_{\mathrm{+,1.0}}$ is the number of stars 1 mag below the tip. The Tip Color is in mag from F606W-F814W.  The Milky Way extinction is from \cite{Schlafly11} and the Internal Extinction is calculated from \cite{Menard10}.  }
\end{deluxetable*}

\begin{deluxetable}{ |c c c c|} 
\tablecaption{Summary of SN Ia in TRGB Hosts}
\tablehead{\colhead{SN} & \colhead{Host} & \colhead{$m_B^0$ (error)} & \colhead{\# LC}} 
\startdata
       2011fe & M101 &  9.808 $\pm$  0.116  &            2\\
      1989B & M66 &  10.980 $\pm$  0.150  &            1\\
      1998bu & M96 &      11.000 $\pm$  0.150  &            1\\
   1980N & NGC 1316 &  12.002 $\pm$  0.097  &            1\\
      2006dd & NGC 1316 &  11.940 $\pm$  0.108  &            1\\
      1981D & NGC 1316 &  11.610 $\pm$  0.230  &            1\\
       2012fr & NGC 1365 &  11.915 $\pm$  0.119  &            2\\
 2011iv & NGC 1404 &  11.974 $\pm$  0.099  &            1\\
       2007on & NGC 1404 &  12.460 $\pm$  0.190  &            1\\
2001el & NGC 1448 &  12.254 $\pm$  0.136  &            1\\
       2021pit & NGC 1448 &  11.752 $\pm$  0.200  &            1\\
 
2007sr & NGC 4038 &  12.434 $\pm$  0.112  &            2\\

2012cg & NGC 4424 &  11.496 $\pm$  0.206  &            2\\
         1994D & NGC 4526 &  11.532 $\pm$  0.093  &            1\\
1981B & NGC 4536 &  11.551 $\pm$  0.133  &            1\\
          2013aa & NGC 5643 &  11.290 $\pm$  0.102  &            2\\
      2017cbv & NGC 5643 &  11.265 $\pm$  0.079  &            2\\
 2021rhu & NGC 7814 &  11.920 $\pm$  0.150  &            1\\
\hline
     
\enddata
\tablecomments{The standardized SN brightnesses from Pantheon+ for each SN used in this analysis. The $\#$LC can be greater than 1 because the SN was observed by more than one survey.} 
\label{tab:snset}
\end{deluxetable}

\subsection{Type Ia supernova Measurements from Pantheon$+$}

For our distance ladder measurement, we use standardized SNe Ia in the second and third rungs.  We make use of the compilation of redshifts and distances from the Pantheon$+$ sample as described in \cite{Scolnic22} and analyzed for cosmological purposes in \cite{Brout22a}. 
A similar study by F19 used a mixture of literature-based SN Ia data for SNe in TRGB hosts (rung 2) and one specific survey only for SNe in the Hubble flow or 3rd rung, the CSP sample. The need to combine disparate sources of SNe for rung 2 is unavoidable; SNe in this volume are rare and their collection occurs over decades and many survey lifetimes.  However, different sources of SN data may have differences in their photometric calibration that can be removed through the use of all-sky survey stellar catalogues and SN sample comparisons \cite{Scolnic15,Brout22b}.  Even without such SN survey standardization, the presence of SNe from the same survey in both rung 2 and 3 will help cancel survey errors \citep{Brownsberger22} and is thus desirable.  

Pantheon$+$ and other widely used compilations, such as JLA \citep{Betoule14}, employ corrections to host redshifts (starting from the CMB restframe) to improve their convergence to the Hubble flow.  These are based on well-sampled maps of the local flow (i.e., peculiar velocity) constructed from 2MASS.  Pantheon$+$ also reassigns the host redshift to that of its host group or cluster when applicable, which further reduces the variance from the Hubble flow.  Both steps have been shown to markedly improve the dispersion of the SN Ia Hubble diagram, while lowering systematic errors in $H_0$ and the equation of state of Dark Energy due to coherent flows \citep{Carr22,Peterson22}.

It also common practice to apply quality cuts on the SN sample \citep{Scolnic18} to limit systematic uncertainties due to the pull of ``extreme'' SNe (in color or light curve shape) that are far from the middle of the distribution and thus whose standardization is likely less accurate.  The most common selection is $-3 < x_1 < 3$ and $-0.3 < c < 0.3$ \citep{Scolnic18,Scolnic22,Betoule14}.  Tighter limits empirically yield lower SN Ia dispersion on the Hubble diagram, and are preferable when the size of the sample does not limit the statistics of the measurement; for example, the SH0ES analysis (R22) imposed tighter limits of $-2 < x_1 < 2$ and $-0.15 < c < 0.15$, and a number of the SNe in TRGB hosts would not pass these (1989B and 1998bu have $c \sim 0.3$, 1981D has $c \sim 0.2$, and 2011iv and 2007on have $x1 \sim -2$).  As discussed in \cite{Rose22}, some SNe that fall out of the color cut are still quite well fitted by the SALT2 model and yield valuable distance measurements, although a study of SN Ia with abnormal shape has not been done.  To avoid limiting the already small TRGB host sample and to better match literature studies, we adopt the broader selection, which retains the two reddened SNe, 1998bu and 1989B, and the faster decliner, 2007on.  In keeping with our unsupervised approach, we retain all SNe within these bounds. We do not exclude \textit{a priori} any tip measurements based on the inferred host distance; we will, however, exclude outliers at the $ > 3~\sigma $ level in the subsequent \textit{global} fit of the relation between SN magnitude and TRGB (see \S~3.2.2).  \footnote{\cite{Freedman19b} included SN2007on in NGC 1404 as did \cite{Anand22}, though \cite{Freedman21} subsequently excludes it citing a $\sim$ 0.4 mag difference with its sibling in the same host, SN2011iv, as well as its difference with SNe in NGC 1316, a galaxy which they presume is at a similar distance as NGC 1404.   In our baseline analysis we find SN2007on to be 1.9$\sigma$ from the global SN-tip relation, hence it is not excluded.  We also note two other siblings in the TRGB host sample, 2021pit and 2001el, both in NGC 1448, with a similar sized difference, so sibling rivalry at this level is not without precedent \citep{Scolnic20}.}

Besides the SNe in the Pantheon$+$ set, we also include ZTF SN~Ia SN 2021rhu which recently was discovered in NGC 7814 \citep{Dhawan22}. While the ZTF photometric system was not calibrated to PS1 in \cite{Brout22b} due to its recent emergence, here we measured the mean difference of the ZTF Hubble flow and Pantheon$+$ sample.  To do this, we utilize the public data release in \cite{Dhawan22} and with the same SALT2 cuts as Pantheon$+$, we find agreement at the $0.02\pm0.01$ mag level, showing that there are no expected systematics that rival the statistical errors for including one SN measured by ZTF \citep{Brownsberger22}.   

In Table \ref{tab:snset} we provide the mean standardized SN Ia magnitudes (denoted $m_B^0$) and their uncertainties for all of the SNe in the TRGB hosts.  These magnitudes will be used to tie the TRGB brightnesses to the SN brightnesses in the second rung of the distance ladder, as discussed in \S 3.  We place an asterisk next to the names of the SNe that in \S 3 are found to lack a TRGB detection, as discussed in the Appendix.

\section{Method and Nominal Results}

\subsection{TRGB EDR Standardization}

\begin{figure}
    \centering

    \includegraphics[width=0.5\textwidth]{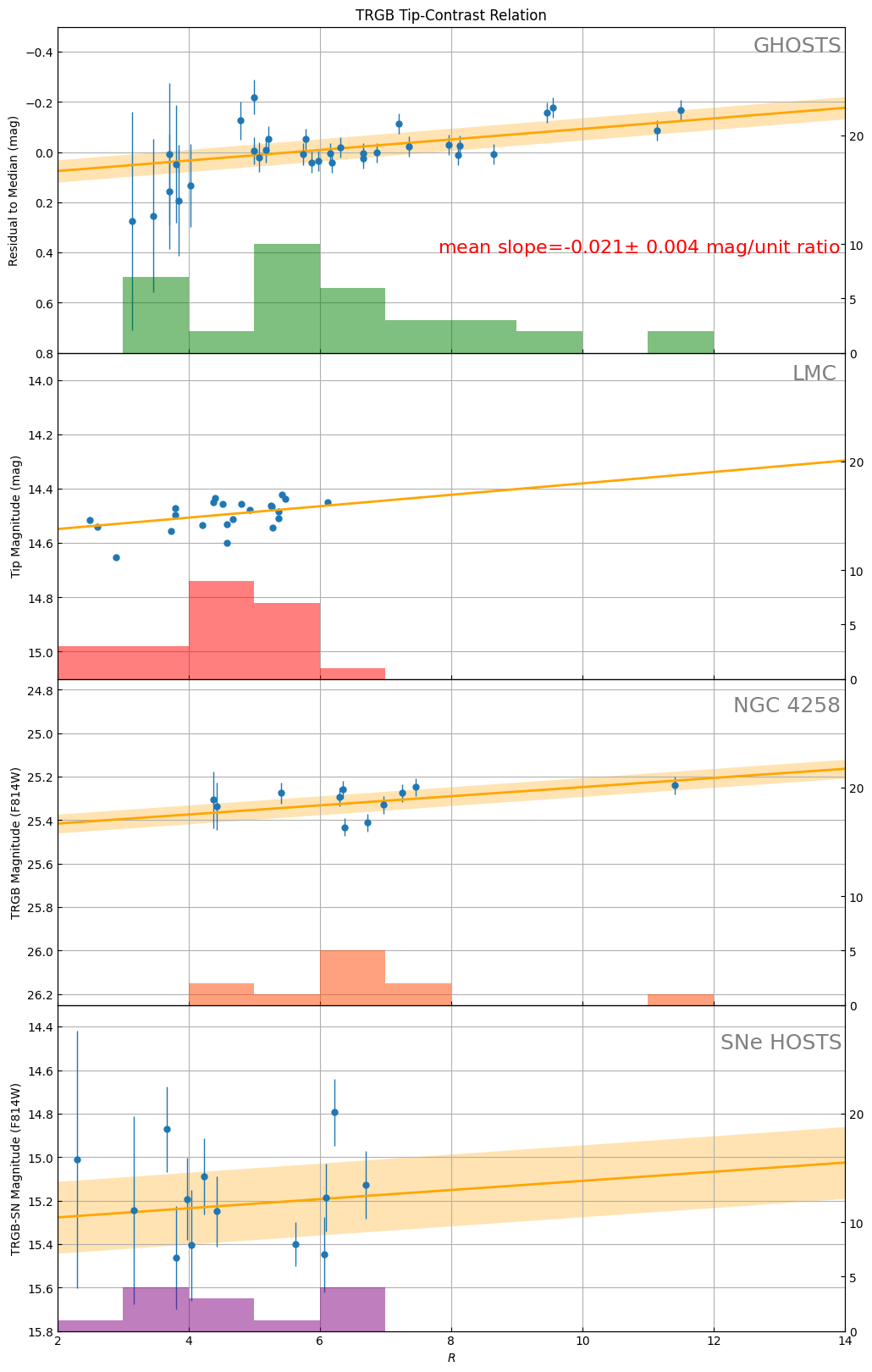}
    \caption{TCR relation for GHOSTS galaxies (top), LMC (second) NGC 4258 (third) and SN host galaxies (bottom). We do not include uncertainties for the LMC points as the values are taken from \cite{Hoyt23} and did not apply the same CATs algorithm for measuring tip brightness.
    \label{fig:trends}}
\end{figure}

In W22, we found evidence at the $5\sigma$ level of an empirical dependence of the measured tip brightness (i.e., the TRGB) and the previously defined contrast ratio, $R$.
L23 independently finds a slope of $-0.015 \pm 0.008$ using 11 fields of NGC 4258, in good agreement with the slope found in W22 of $-0.023 \pm 0.0046$.  Combined, we find a slope of $-0.021\pm0.004$, which we overlay on the data samples shown in Fig.~\ref{fig:trends}. We show that a similar relation also may apply to the 25 ranked LMC fields analyzed by \cite{Hoyt23}, where we use the tip values from \cite{Hoyt23} and calculate the $R$ values from the catalog and regions provided.  The $R$ found here correlates to some extent with the field rank discussed in \cite{Hoyt23}, which was shown in that study to correlate with tip brightness. As shown in L23, there is also theoretical support for this relation; if CMDs extracted from MIST isochrones \citep{mist16} using different ages and metallicities are subjected to the same noise and measurement procedure at the level of half to the full empirical level, the resulting tip brightnesses correlate with the ratio of RGB to AGB stars.  Some of this relation may be the result of the measurement process itself: the more asymmetric the population near the break, and thus the higher the contrast ratio, the more stars will ``spill'' to the bright side of the break, thereby increasing its brightness. 
This was also shown by \cite{Anderson23}, both by simulating luminosity functions and by empirically different TRGB magnitudes resulting from two individual sequences of variables that otherwise blend into the RGB.
Regardless of whether this relation is driven by astrophysical or measurement effects, an accurate measurement of $H_0$ requires the standardization of the tip brightness between rungs of the distance ladder.  Therefore, to avoid a bias which would occur if the mean value of the contrast, $R$, is different in SN hosts than in NGC 4258, we use the contrast ratio to rectify the tips.  

Since this relation is consistent between the NGC 4258 and the GHOSTS sample (W22), we apply their mean linear correction for the TCR to remove the slope dependence.  We must choose a fiducial (but arbitrary) $R$ value to correct to a standard tip.  We choose the value of $R=4$ because it is near the middle of the SN host values (median $R=4.2$, mean $R=4.5$), making it a useful reference calibration.  

Therefore, we can derive a standardized tip magnitude of each field measurement $m_{I,TRGB}^{R=4}$ using the measured, `raw' tip magnitude $m_{I,TRGB}$ such that

\begin{equation}
m_{I,TRGB}^{R=4}=m_{I,TRGB}-0.021(R-4).
\label{eqn:correct}
\end{equation}
We note that the reference value of $R=4$ has no effect on the determination of $H_0$, as long as the value is the same for the correction in the first and second rung.  In \S 4 we will also provide results without use of the TCR or $R$ in general to determine its impact on $H_0$. Finally, we use the empirically-measured tip uncertainty model produced in W22 which, depending on the $R$ and $\Nminusone$ are the specific values for the field, which is:

\begin{equation}
 \sigma=\sqrt{\left[\left(\frac{2e^{1.5(3-R)}}{e^{1.5(3-R)}+1}\right)\left(\frac{1}{\Nminusone-100}\right)^{0.1}\right]^2+0.04^2}\;\mathrm{mag}
 \label{eqn:error}
\end{equation}

Plots showing the dependence of the uncertainty on $R$ and $N_{+,1.0}$ are given in W22. As the formula shows, there is a dramatic loss in precision for very fuzzy (low contrast) tips with $R<3$ and those with $R<2$ having effectively no value. We therefore exclude from consideration tips with $R<2$ as either spurious or of no weight.

We find that the four most distant hosts, NGC 5584, NGC 1309, NGC 3370 and NGC 3021 (based on their SNe) do not yield a tip with $R>2$ {\it at any location within a magnitude of the expectation from their SNe}, including the magnitude indicated by \cite{Jang17}. As discussed in Appendix B, this is a strong empirical qualification which is independent of the specific detection methodology; due to its low contrast, the uncertainty in the tip magnitude, at any plausible magnitude, is too low for it to carry any significant weight.  This is not surprising, as given observations of these four galaxies, there is a very small useful region for a clean TRGB measurement; the original observations targeted the disk in order to measure Cepheids, leaving very little area free from disk contamination. This is consistent with the finding by \cite{Anand22}, who also found no meaningful break in the LF of these hosts using LF fitting.  We cannot readily attribute this difference with \cite{Jang17} to a difference in depth due to the photometry tools (we use DOLPHOT, \citealp{Jang17} used DAOphot) because 1) they are based on the same images, 2) they both use image stacks to produce a full source list and 3) the depth of the DOLPHOT photometry matches the expectation based on the {\it HST} exposure time calculator.  Furthermore, the photometry catalog of \cite{Jang17} is not available, so further understanding this difference cannot be done.

\subsection{Determining $H_0$}

The determination of $H_0$ from a three-rung ladder (connecting the geometric distance to NGC 4258, standardized TRGB and standardized SN Ia) can be succinctly determined by measuring a single quantity for each rung independent of the others.  While distance-ladder analyses like \cite{Riess22} measured covariances between the three rungs, the approach we are taking by measuring each rung separately favors transparency and ease of comparison to other studies which we think is paramount for the TRGB method at this juncture.  There is a very small loss of information by forgoing the use of non-diagonal covariance of SNe between rung 2 and 3 and the simultaneous optimization of the TCR between rung 1 and 2 with the present use of the GHOSTS sample (W22).

\subsubsection{First Rung}
The first rung entails the geometric calibration of the TRGB, 

\begin{equation}
\textbf{\textrm{~~}} M_{I,TRGB}^{R=4}=m_{I,N4258,TRGB}^{R=4}-\mu_{0,N4258}
\label{eqn:firstrung}
\end{equation}

where $\mu_{0,N4258}$ is the geometric distance to NGC 4258
and $m_{I,N4258,TRGB}^{R=4}$ is the apparent TRGB in NGC 4258 in the HST ACS $F814W$ bandpass, rectified to a fiducial contrast ratio of $R=4$. 

The geometric measurement of $\mu_{N4258}=29.397\pm0.0324$ mag is from \cite{Pesce20}, and the value of $m_{I,N4258,TRGB}^{R=4}$=\ghostcal~mag is from L23.
We note that the 11 halo or partial halo fields in NGC 4258 analyzed by L23 have individual apparent tips with meaningful contrast ratios from $4 < R < 11$  with a substantial range of $m_I=25.6$ to 25.3 mag. 
The range in both contrast ratio and tip magnitude, \textit{and their correlation,} illustrates the need for standardization to a common reference $R$.  This avoids a subjective selection of a subset of tip values which may not match the properties of tip measurements in SN hosts.  The value we adopt from L23 includes a small mean correction for extinction within NGC 4258 (following \citealp{Menard10}) of $A_I=0.013$ 
mag and a mean correction for MW extinction of $A_I=0.029$; for comparison with literature values without any extinction correction or TCR standardization, e.g., \cite{Jang21}, the uncorrected mean tip magnitude of all tips in all fields in NGC 4258 is \simpleavecal. Alternatively, weighting the simple, uncorrected tips by the number of stars below the tip yields \numsimpleavecal.  We then calculate a fiducial tip luminosity of $M_{I,TRGB}^{R=4}=-4.030 \pm 0.035$ mag, or $M_{I,TRGB}^{R=4}=-4.018 \pm 0.035$ 
without internal extinction.  For comparison to studies without internal extinction correction or standardization (i.e., contrast weighting), i.e., a simple average of all NGC 4258 fields correcting only for MW extinction, this would be $M_{F814W}=-4.026 \pm 0.050$ mag (though at a mean $R$ of $\sim$ 6.6).  Alternatively, if we treat the average of the tips as derived from a superposition of fields, i.e., as though it was a more distant host with a single field, we weight the tips by the number of stars below the tip which yields an average corrected for MW extinction of $M_{F814W}=-3.956 \pm 0.050$.

\subsubsection{Second Rung}

The second rung can be measured from the mean of a single apparent quantity, the difference between the standardized TRGB, $m_{I,TRGB}^{R=4}$ and the standardized SN magnitude $m_B^0$ in the same galaxy.  If a galaxy has multiple SNe, we use their average magnitude $\overline{m_B^0}$.  Because this term is the difference of two standardized candles in the same host, it is expected to be a constant across the sample.  We will call this quantity $\Delta S$:  

\begin{equation}
\textbf{\textrm{~~}} \Delta S=m_{I,TRGB}^{R=4}-\overline{m_B^0}
\label{eqn:secondrung}
\end{equation}

Then $\overline{\Delta S}$ is the value of $\Delta S$ averaged over the individual hosts, with weights that are the sum of the inverse error squared for each component. The values of $m_B^0$ are taken from Pantheon$+$, as discussed in \S 2. In cases where there are two SNe per host, we take the weighted mean and its error, and for three SNe per host we take the error as their internal dispersion divided by $\sqrt{N-1}$. 

The GHOSTS survey sample used to develop the CATs algorithm had the benefit of multiple fields per host, so we could exclude ``outlier tips'' which were too far from the host mean.  The same is possible in NGC 4258 with its 11 fields (L23).  Past literature studies have referred to specific TRGB measurements as being ``unsuitable'' on a case-by-case basis, relying either on subjective factors such as appearance, or on external information, such as an HI map for contamination from a young population, which may not be uniformly available along the distance ladder.   Nevertheless, it may be worthwhile to exclude some TRGB or SN measurement that appear anomalous in value.  For the second rung, $\Delta S$ can serve this function, because it has a constant expectation value for all hosts, so it can be used to identify true outliers. We adopt a $3\sigma$ ($\chi^2 > 9$) cut in the computation of the weighted mean $\overline{\Delta S}$.  Using our baseline algorithm optimized via GHOSTS, we find that none of the host-SN combinations and none of the NGC 4258 fields fall into the outlier category with this definition; but we will make use of this form of outlier rejection when we consider variants of the CATs algorithm in \S 4. For our baseline analysis, we measure $\overline{\Delta S}=$\ghostsn ~mag (i.e., the fiducial SN Ia is that many magnitudes brighter than the fiducial TRGB).

The combination of equations \ref{eqn:firstrung} and \ref{eqn:secondrung} (i.e., rungs one and two) provide the luminosity calibration of standardized SN Ia, 

\begin{equation}
M_B^0=M_{I,TRGB}^{R=4}-\overline{\Delta S}
\end{equation}

\noindent from which we obtain $M_B^0=-19.245 \pm 0.060$ mag.  This quantity is directly comparable to $M_B^0=-19.269 $ from the SH0ES measurement using Cepheids in NGC 4258 (as the only anchor) and 42 SNe Ia in 37 Cepheid-SN hosts (R22).  These values are in excellent agreement, with a difference of 0.024 mag; the independent uncertainties (excluding SNe in common) add up in quadrature to an uncertainty of $\sim$ 0.050 mag.  The baseline SH0ES result using 3 independent geometric anchors is $M_B^0=-19.253 \pm 0.027$, which is also very close to the TRGB result.

\subsubsection{Third Rung}

The third rung of the distance ladder is derived from SNe Ia only and is thus separable from any TRGB measurements. It can be summarized by the intercept of the Hubble diagram, $a_B=\log\,cz - 0.2m_B^0$ in the low-redshift limit ($z \approx 0$), where $z$ is the redshift, after corrected for peculiar velocities \citep{Carr22,Peterson22}. $m_B^0$ is the maximum-light apparent magnitude of the SNe which has been {\it standardized} (i.e., corrected for variations around the fiducial color, luminosity, and any host dependence following Pantheon$+$).  For an arbitrary expansion history, $ a_B $ is expressed at small redshift by:
\begin{equation}
\begin{split}
& a_B=\log\,cz \{ 1 + {\frac{1}{2}}\left[1-q_0\right] {z} \\
 & -{\frac{1}{6}}\left[1-q_0-3q_0^2+j_0 \right] z^2 + O(z^3) \} - 0.2m_B^0 \label{eq:aB} 
\end{split}
 \end{equation}

Here $q_0$ is the deceleration parameter, and $j_0$ is the jerk parameter \citep[see][for definitions]{Visser04}.  We use a value of

\begin{equation}
a_B=0.71448 \pm 0.0012
    \end{equation} 

\noindent derived in R22 (and $q_0=-0.55$ and $j_0=1$) for the ``all host types'' at $0.0233 < z < 0.15$, from 484 SNe Ia, appropriate for a TRGB sample that includes early type hosts (e.g., NGC 1404 and NGC 1316).  We discuss this value further in \S 5.

We note that different surveys providing SN Ia magnitudes may yield different results for individual SNe, but these differences largely cancel provided that the sample of $m_B^0$ are consistently calibrated using an all-sky survey \citep{Brout22a} and SNe from these same surveys are also used to measure the Hubble intercept, $a_B$ \citep{Brownsberger22}.

\subsubsection{Baseline $H_0$}

Combining the constraints from all three rungs, the Hubble constant $H_0$ is determined as

\begin{equation}
 \log{H_0}={0.2 M_B^0\!+\!a_B\!+\!5}. \label{eq:h0alt} 
\end{equation}

\noindent from which we find $H_0$=\ghostH.
 
The uncertainty (in magnitudes) in the above follows from the quadrature sum of the above errors in $\mu_{N4258}$, $m_{I,N4258,TRGB}^{R=4}$, $\overline{dS}$, and $5a_B$:

\begin{equation}
\sigma_{top} = \frac{\sqrt{\sigma^2_{\mu_{0,N 4258}}+\sigma^2_{{m^R_{N4258}}}+\sigma^2_{\Delta S}+\sigma^2_{5a_B}}}{2.17} \times H_0
\label{eqn:uncert}
\end{equation}

In magnitudes (km/s/Mpc), we find that the contributions from the four terms are 0.032 mag (1.1 km/s/Mpc); 0.014 mag (0.5 km/s/Mpc); 0.050 mag (1.7 km/s/Mpc), and 0.006 mag (0.2 km/s/Mpc) respectively. The quadrature sum is 0.061 mag or 2.8\% or 2.06 km/s/Mpc.

These results represent the application the TRGB algorithm as optimized to produce the lowest inter-host variance of the tip for the GHOSTS training sample.  However, there are uncertainties in the algorithm parameters which are difficult to propagate in any analytical form.  Therefore, in the next section we consider variants of the algorithm parameters, from which we can determine other representative values of $H_0$ and an algorithmic systematic uncertainty.

\begin{deluxetable*}{| c | c c c |} 
\tablecaption{Baseline and Representative Results on $H_0$}

\tablehead{\colhead{Method} & \colhead{$H_0$}    &   \colhead{$m_{I,N4258,TRGB}^{R=4}$} &  \colhead{$\overline{\Delta S}$}}  
\startdata                           
GHOSTS Baseline & \ghostH & \ghostcal & \ghostsn \\
Best SN Dispersion  &    \lowdispH &  \lowdispcal & \lowdispsn \\
Best $\chi^2$ & \lowchiH & \lowchical & \lowchisn \\
Median $H_0$   & \medianH & \mediancal & \mediansn \\
\enddata

  \tablecomments{The recovered values for specific variant cases which produce optimal results.  The derivations for 
  $H_0$, $m_{I,N4258,TRGB}^{R=4}$, $\overline{\Delta S}$ are given in Eq.~\ref{eq:h0alt}, Eq.~\ref{eqn:firstrung} and Eq.~\ref{eqn:secondrung} respectively. The results are discussed in \S \ref{subsec:changingparam}. Each variant includes all SN fields and NGC 4258 fields with tip measurements.}
     \label{tab:representative}
  \end{deluxetable*}

\section{Variants of TRGB Measurement Algorithm}

\subsection{Changing algorithm parameters}
\label{subsec:changingparam}

In \S 2.1 we described four features of a general TRGB measurement algorithm, each controlled by a parameter; we determined the preferred parameter values as those that produce the lowest dispersion among field tips measured around halos of the same hosts using the GHOSTS sample (W22) with a distance range of 2-15 Mpc (and which included one of the SN hosts studied here, NGC 7814).  Because of the overlap in distances and contrast ratios between the GHOSTS sample and the SN hosts and fields in NGC 4258, we expect those choices to be appropriate for our present analysis.  However, in order to understand the sensitivity of our results to those parameter choices, here we consider a wide range of variations which bracket their baseline values, and thus estimate a systematic uncertainty in the determination of $H_0$ due to algorithm choices.

These variants are:
\begin{itemize}

\item $\textit{\textrm{SpatialClip~\%}}$ is varied from the nominal value of $>10\%$ with values of $>5, >10, >20\%$.

\item  $\textit{\textrm{Width}}$ of the color band for the CMD is varied from the nominal at 1.0 mag to $0.75, 1.0, 1.5, 2.0$ mag.  Here we included a fourth, very large (near limitless) width, to produce a more relevant comparison with the CCHP analysis in F19 for which little to no limitation in width was applied (T. Hoyt, private communication).

\item $\textit{\textrm{Smooth}}~\sigma$ of the luminosity function is varied around the nominal of $\sigma=0.1$ with three options, namely $\sigma=0.07, 0.1, 0.15$.

\item $\textit{\textrm{MinTh~\%}}$ is varied from the nominal value of $0.6\times$ the highest EDR value, with values of $0.5$, $0.6$, and $0.8$.

\end{itemize}

\noindent The above choices provide a total of $3\times4\times3\times3=108$ variations on the analysis. We can characterize the variation in $H_0$ they produce in order to determine a systematic uncertainty relating to the TRGB algorithm.  The variants also may be considered with appropriate context, to identify some interesting and perhaps representative determinations of $H_0$ such as the median of all variants, the variant with the lowest $\chi^2$ or dispersion, or that most similar to some literature analyses.

\begin{figure*}
    \centering
\includegraphics[width=1.0\textwidth]{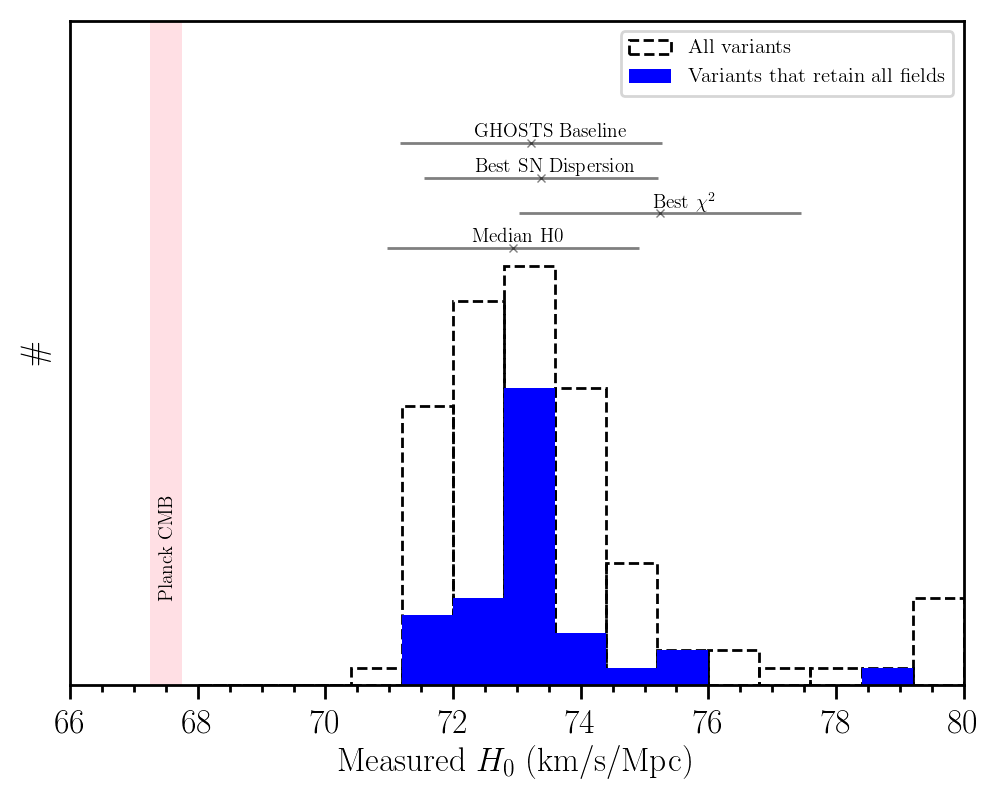}
    \caption{The distribution of recovered $H_0$ values for the variants described in \S \ref{subsec:changingparam}.  The values for the full set of 108 variants is shown in the black outline.  The filled blue shows only the variants in which tip measurements of all the SN fields and all the NGC 4258 fields are recovered.  We show the constraint on $H_0$ from \textit{Planck} as well (pink). The horizontal lines around $H_0=73$ show the values recovered in Table~\ref{tab:representative}. }
    \label{fig:summresults}
\end{figure*}

We summarize each variant in a table in the Appendix with the first four columns describing the analysis parameters, followed by the determination $H_0$, its error, the values of the first and second rung parameters, the total reduced $\chi^2$ of the first and second rungs, the dispersion $\sigma$ in the second rung, the number of TRGB tips found for the SN hosts, the number of SN hosts with a valid tip (a maximum possible of 13), and the total number of SNe contained in these hosts (a maximum possible of 18).  We show the distribution of the measurements of $H_0$ in Fig.~\ref{fig:summresults}. The formal dispersion of all variants is 2.94 km/s/Mpc, which exceeds the statistical error of 2.05 km/s/Mpc.  There is a tighter core of values and a long, low tail towards high $H_0$ values.  While our baseline measurement found no outliers among the SN or NGC 4258 fields, for a minority of variants one or more fields are excluded as $>3 \sigma$ outliers. To better understand how much of the variation in $H_0$ is due to the variation in the algorithm, we limit consideration of the variants to those which retain all SN hosts and NGC 4258 fields, some 55\% of the variants. Doing so also largely eliminates the high $H_0$ tail as seen in Fig.~\ref{fig:summresults}, and the remaining 59 variants have a median of {\medianH}and a standard deviation of $\sim 1.33$ km/s/Mpc.  We consider this spread a more fair representation of the systematic uncertainty due to the analysis process, as it does not include variations due to the exclusion of some data points. The full range of these variants is 71.2 to 78.4.  

\begin{figure}
    \centering
    \includegraphics[width=0.5\textwidth]{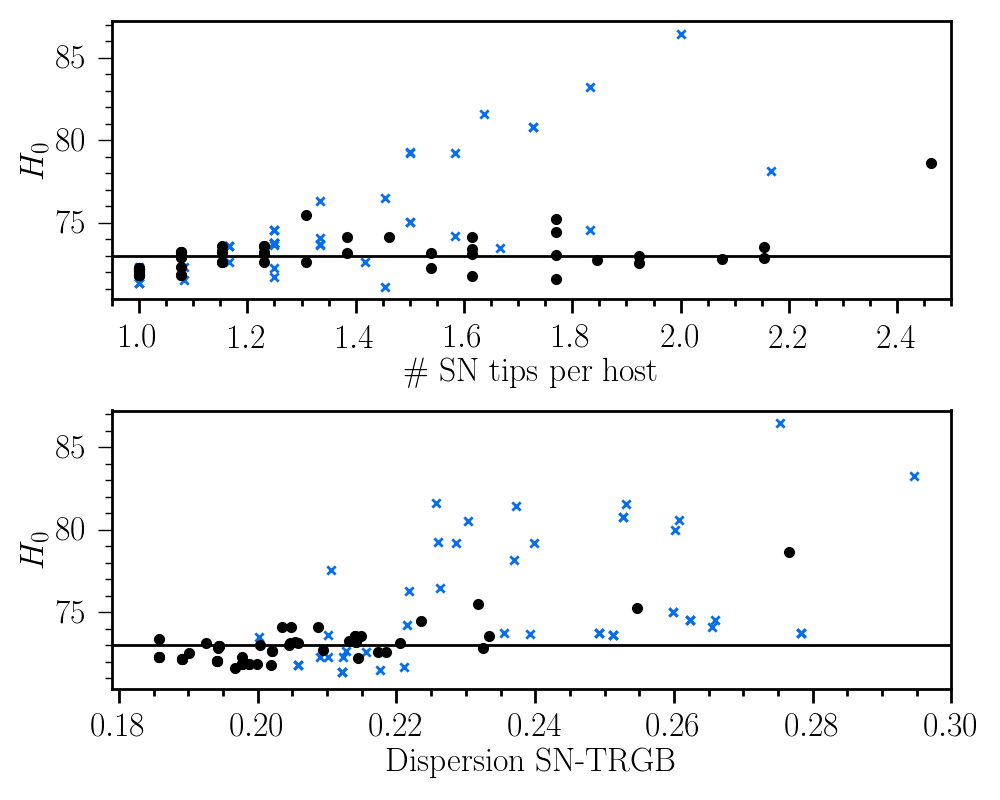}
    \caption{Inferred value of $ H_0 $ for all the analysis variants, as a function of the average number of tips per host (top) and of the dispersion of the difference between SN and TRGB distances(bottom).  Horizontal lines are drawn at $H_0=73$.}
    \label{fig:H0trends}
\end{figure}

To further study the high $H_0$ tail seen in Fig.~\ref{fig:summresults}, we show in Fig.~\ref{fig:H0trends}
that the source of the high $H_0$ tail can be seen as arising from variants that produce excess, i.e., ``spurious'' tips.  Ideally, each host should have one tip of the RGB.  In practice, application of the EDR to hosts with a combination of poor contrast (low $R$), significant photometry noise, and a finite number of stars can in some cases produce multiple local maxima, all of which apply here.  This consequence appears unavoidable. Following the approach we defined in Paper I (W22), we retain multiple peaks, weight them by a function of their contrast ratio, and subject them to formal outlier clipping.

Similar to what was seen in W22, for variants where the number of tips per field approaches unity ($<1.4$), the results become more stable. The variants with a large ratio of tips per field tend to be those with lower smoothing, higher spatial clipping, and a lower minimum threshold.  Overall, 66 of 108 variants have fewer than 1.4 tips per host, and 42 have more.  We find that the largest driver for these is smoothing; the mean smoothing for variants with a tip-to-field ratio $<1.4$ is 0.125, while it is 0.078 for variants with tip-to-field ratio $ > 1.4 $.  Interestingly, the smallest scatter for NGC 4258 is found with less smoothing; for smoothing of 0.07, 0.10, 0.15 mag, the dispersion in tip magnitude is 0.047, 0.055, 0.054 mag respectively, including a lowest scatter of 0.027 mag for the case with 0.07 mag smoothing, 2.0 mag color width, and 20\% spatial clipping threshold.  It is also worth nothing that the high $H_0$ tail is supported by variants with less spatial clipping (the $>$20\% threshold) with a modest correlation between $H_0$ and the level of spatial clipping, as seen in Figure \ref{fig:summresults}.  

\begin{figure*}
    \centering
  
    \includegraphics[width=1.0\textwidth]{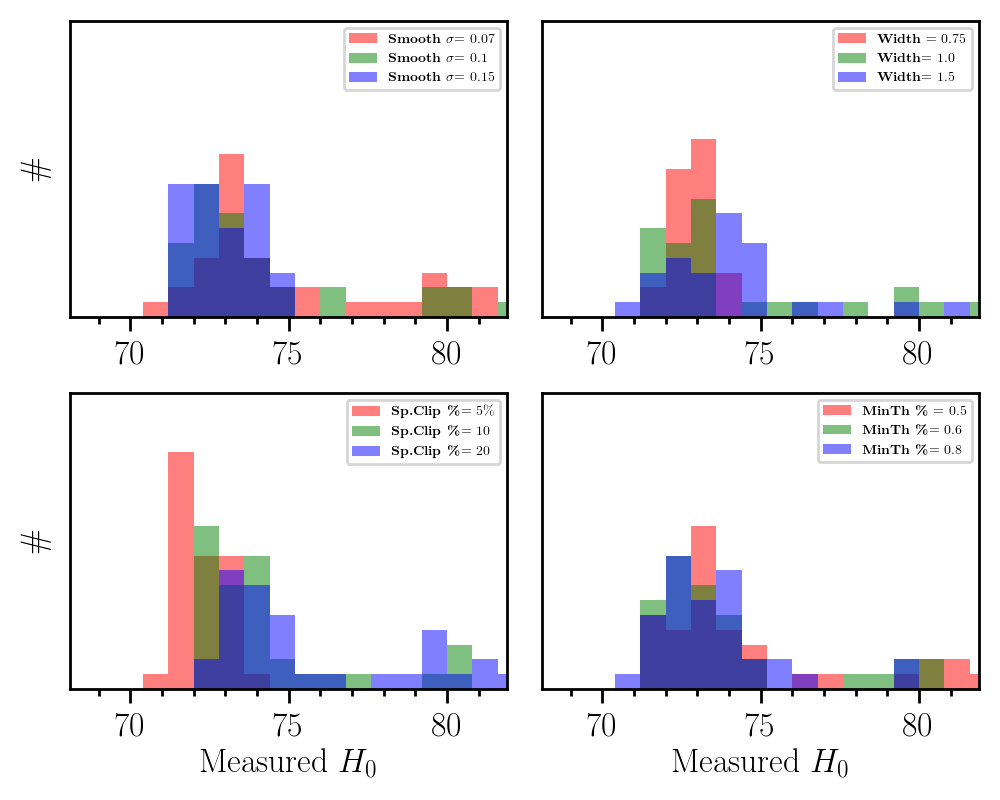}
    \caption{The impact of different variants in the analysis, summarizing the output of Table~ \ref{tab:variants}. (Top left) Recovered values of $H_0$ for all variants, broken into subsets according to smoothing values used. (Top right) Same, but broken into subsets using the widths of the color bands. (Bottom left) Same, but broken into subsets using the fraction of the field clipped spatially.  (Bottom right) Same, but broken into categories varying the minimum threshold.   }
    \label{fig:H0variant}
\end{figure*}

In the four panels of Fig.~\ref{fig:H0variant} we examine the impact of the different analysis parameters on $H_0$.  The only obvious pull can be seen for the spatial clipping $\%$s, where the mean $H_0$ of the $5\%$ distribution is $\sim0.7$ km/s/Mpc lower than for $10\%$ or $20\%$. This can be traced to two fields of NGC 4258 where the more aggressive spatial clipping of 5\% shifts the inferred tip value $\sim0.05$ brighter.   Some of the other variation in $H_0$ seen in Fig.~\ref{fig:H0variant} can be traced to $\sigma$ clipping.  The most common SN field to be $3 \sigma$ clipped from the SN host sample is NGC 4038, which is eliminated in 40 of 108 variants, a likely consequence of its greatest distance and fuzzy, ambiguous EDR with two strongest and distinct peaks.  NGC 4526 fails in 13 of 108 variants.  The other fields rarely fall out.

Based on the variants, we can define additional results of interest by ``reoptimizing'' the algorithm parameters with different parameter values, aiming to produce the either the lowest dispersion between the SNe and tips or the lowest $\chi^2$ overall.  Doing so, these variants give \lowdispH~with $\sigma=$ \lowdisp~mag and \lowchiH~with $\chi^2_{dof}$=\lowchi, respectively.  We summarize these results as `Best SN Dispersion' and `Best $\chi^2$' in Table~\ref{tab:representative} respectively.


\subsection{Effect of Tip-Contrast Standardization on $H_0$}
\label{subsec:notcr}

In this subsection we consider the impact of ignoring the contrast ratio in our TRGB measurements: thus no contrast ratio weighting of tip measurements, and no standardization of the tip brightness. Determining these numbers facilitates comparison to literature studies.  We summarize possible results when neglecting tip standardization in Table~\ref{tab:neglect}.

To first approximation, the impact of the tip-contrast relation on $H_0$ results from the difference in mean contrast, $R$, for fields around NGC 4258 and SN hosts. We measure a weighted mean $R$ (using Eq.~\ref{eqn:error} as the weights) for NGC 4258 of 7.2 and for the SN hosts of 5.2.  This $\Delta R\sim2$ based on Eq.~\ref{eqn:correct}, biases the luminosity calibration of SNe from NGC 4258 by $0.021\times2=0.042$ mag, which is roughly 1.6 units in $H_0$.  The simple average values of R (i.e., unweighted) are smaller at 6.6 and 4.6, respectively, but they have the same difference.  Hence we might expect $H_0$ to be $73-1.6 \sim 71.4$ by ignoring the tip-contrast relation (i.e., TCR=0) or an equivalent method of tip standardization.  However, in practice, estimating $H_0$ without use of the contrast ratio is more complex due to its role in distinguishing between low and high quality peaks in the same field. 

The use of the contrast ratio, $R$, in W22 helped solve the common ambiguity of more than one strong peak in the EDR function used to locate the TRGB.  As seen in Eq.~\ref{eqn:error}, the dependence of the tip uncertainty on $R$ is quite steep, so this formula effectively selects the higher $R$ tip by giving it far greater weight.  For example, the two tips seen in Table \ref{tab:TRGBresults} in NGC 1404 have $R=6.9$ and $R=4.0$, producing a 25 to 1 weighting.  Without the use of the TCR and $R$-weighting, we need to define a universal algorithm to measure the TRGB in the presence of multiple tips.

We tried several approaches, including selecting the highest peak in the EDR (called `Highest EDR Peak') or the peak with the highest value of $R$ (called `Highest $R$ Peak') when there is more than one in a field.  Alternatively, we can take a straight average of the tips of multiple peaks (called 'Average of Peaks').  Lastly, there are a number of algorithmic variants that only produce a single peak for each field (called 'One-tip variants'); for example, using a high smoothing value, $ s=0.15$, acts to merge smaller peaks into one.  
We summarize the results in Table~\ref{tab:neglect} of all of these options.  Methods that favor the highest EDR peak (by smoothing or selecting) yield $H_0 \sim 71.7$, consistent with the first approximation of ignoring the TCR and the use of $R$ to select the TRGB.  Averaging multiple peaks or selecting the one with the highest R results in $H_0 \sim 75$.  We conclude there is a rather large ambiguity in $H_0$ related to the issue of selecting the tip among multiple peaks without the use of a strong, quantitative metric like $R$ that can be applied uniformly.

\begin{deluxetable}{| c | c c c |}
  \label{tab:neglect}
 \tablecaption{Neglecting Tip Standardization--Dependence on Method for Selecting Tips}
\tablehead{
\colhead{Method} & \colhead{$H_0$}    &   \colhead{$m_{I,N4258,TRGB}$} &  \colhead{$\overline{\Delta S}$}} 
\startdata
Highest EDR Peak &  $71.7 \pm 2.6 $ & $25.30 \pm 0.05$ & $15.20 \pm 0.05$ \\
Highest $R$ Peak & $74.1 \pm 2.7 $ &   $25.30 \pm 0.05$ & $15.12 \pm 0.05$ \\
Average of Peaks & $75.3 \pm 2.5 $ &  $25.38 \pm 0.04$  & $15.15 \pm 0.05$ \\
One-tip variants& $71.7 \pm 2.6$ & $25.30 \pm 0.05$ & $15.20 \pm 0.05$ \\
 \enddata
 \tablecomments{Results of various methods when ignoring the usage of $R$ to standardize tips.  Each are discussed in \S \ref{subsec:notcr}. }
  \end{deluxetable}

The closest match of our algorithm to the CCHP analysis \citep{Freedman19b} is to neglect the contrast ratio (including the TCR standardization)
and to select a variant with the highest color width, 2.0, while retaining all SN hosts.  There are only 6 variants which satisfy all of these conditions and they have a mean $H_0$ = 71.8 km/sec/Mpc.  As we show in \S~\ref{sec:Literature}, the larger share of the difference in $H_0$ between \cite{Freedman19b} and this work is not related to the TRGB measurement, but rather to the treatment of the SNe Ia sample.

\section{Discussion: Comparison to previous constraints on $H_0$ with TRGB}
\label{sec:Literature}
There have been four recent determinations of $H_0$ based on TRGB and SNe that have substantial overlap of the sample of host galaxies we used: the CCHP team measured $69.8\pm1.9$ km/s/Mpc \citep{Freedman19b}; the EDD group (which used LF fitting rather than EDR to measure the TRGB) found $71.5\pm1.8$ km/s/Mpc \citep{Anand22}; the RAISINs Team \citep{Jones22} used NIR supernova data (rather than optical) with the CCHP tip measurements and recovered $H_0=76.6\pm2.6$ km/s/Mpc; and finally 
\citep{Dhawan22} used only ZTF supernova data and the CCHP tip method with a single TRGB host and measured 76.9 $\pm$ 6.4 km/s/Mpc.  Here, we trace the differences between the results presented in this paper and these.  The differences related to the first two are itemized in Table \ref{tab:h0differences}.  

\subsection{Differences in $H_0$ due to the Supernova-Data/Analysis Component}

\cite{Anand22} use the Pantheon sample of SNe \citep{Scolnic18}, a beta version of the Pantheon$+$ sample used here, and consistent within $0.5$ km/s/Mpc for application to $H_0$ measurements \citep{Brout22}.  \cite{Freedman19b} performs its own analysis of SNe and uses a literature compilation for SNe in the calibrator hosts and the CSP supernova sample \citep{krisciunas17} for the Hubble flow sample.  As discussed in \cite{Brownsberger22} and \cite{Brout22b}, use of different sources of SNe in the first and second rung can lead to biases in the inferred value of $H_0$ due to SN calibration differences. 
Furthermore, unlike the analysis in Pantheon$+$ or the JLA compilation \citep{Betoule14}, \cite{Freedman19b} does not apply peculiar velocity corrections to the Hubble flow sample, although the evidence of their value is $>4 \sigma$ \citep{Peterson22}.

To estimate the impact of these SN-related differences, we modify the Pantheon$+$ analysis in \cite{Brout22b} and \cite{Riess22}.  If we only include CSP Hubble flow data to measure the intercept of the third rung of the distance ladder, rather than all the surveys contained in the Pantheon$+$ compilation, we measure a shift towards lower $H_0$ of 1.1 km/s/Mpc.  This is consistent with the finding of \cite{Brout22b} (see Figure 6) of an 0.03 $\pm 0.015$ mag difference between the CSP survey and the mean of all SN sources.   Although this offset is not significant, it will nevertheless produce a difference of 1 km/s/Mpc if it is the sole source used to measure the Hubble flow in relation to the second rung which uses all samples. 
 
In addition, if we remove the peculiar velocity correction (for the CSP sample), this shifts the inferred $H_0$ lower by $\sim0.4$ km/s/Mpc, consistent with the findings of \citep{Peterson22}.  This amount is somewhat smaller than that found for the full sample in \cite{Peterson22} of $\sim0.6-0.7$ km/s/Mpc because of the location of CSP SNe in the sky and the redshift range of the SNe ($0.02 < 0.15$) used to measure $H_0$.  Combined, these two distinct effects contribute a shift of $1.5$ km/s/Mpc towards a lower $H_0$ relative to \cite{Freedman19b}.  They do not produce any difference with any of the three other studies; EDD \citep{Anand22} and ZTF \citep{Dhawan22} use flow corrections and matched surveys, while RAISINs \cite{Jones22} uses only CSP SNe, but on both rungs of the distance ladder.  These factors partially explain why these other analyses all yield higher values of $H_0$.
 
Additionally, we include two new calibrator SN Ia from the year 2021, SN 2021rhu from \cite{Dhawan22} and SN 2021pit (\citealp{Swift14}; in NGC 1448 which already included SN 2001el).  We also include SN 2007on, used by \cite{Freedman19b} but excluded by \cite{Freedman21}.  The combined impact of the additional 3 SNe Ia is 0.6 km/s/Mpc relative to \cite{Freedman21}.  EDD already included 2007on but did not include the 2021 SNe, so the relative change from the two additional SNe is 1.3 km/s/Mpc.  
 
The four hosts for which neither we nor EDD could detect a reliable tip, i.e., NGC 5584, 3021, 3370, and 1309, as evaluated by reanalyzing the CCHP result with and without them, cause no change in $ H_0 $ when they are excluded from a reanalysis of the CCHP results, although the uncertainty increases by 10\%.

These effects combined account for a net difference of $2.0$ km/s/Mpc between this study and the CCHP {\it for SNe related terms which are independent of the TRGB}.  Simply revising the SN data in \cite{Freedman19b} with Pantheon$+$ would yield $H_0$ of 71.8 km/s/Mpc, which well matches the value we found for the TRGB algorithm variants which most closely match the CCHP procedures. For EDD, the only difference is the two new SNe from 2021 which cause a change of 1.3 km/s/Mpc, and revising the EDD analysis with the new objects would yield $H_0$ of 72.8 km/s/Mpc.  Such changes due to a couple of new SNe is not surprising given the still small number of SNe in well-measured TRGB hosts.

\subsection{Differences in $H_0$ due to the TRGB-Data/Analysis Component}

The remaining differences between our analysis and those in \cite{Anand22} and \cite{Freedman19b} relate to the fiducial calibration of the TRGB derived from NGC 4258 and applied to SN hosts.  As discussed in \S 4, there is a difference in the mean tip contrast between the fields of NGC 4258 and SN hosts of $\Delta R \sim $2 which, via the tip contrast relation of 0.021 mag per unit in $ R $, produces a difference of $+1.6$ km/s/Mpc with this analysis and that from CCHP, though a difference of $-0.3$ km/s/Mpc from EDD due to the specific fields used in that analysis.  

A direct way to understand this difference in $H_0$ derived here versus that derived in CCHP or EDD is by comparing the magnitude of the tip in NGC 4258 estimated by each group. Including removal of MW extinction (and excluding removal of extinction in NGC 4258 for even comparison), the value found here `corrected to the mean $R$ of the SN hosts' is 25.389 $\pm$ 0.00136.  Our simple mean of the NGC 4258 fields (i.e., without use of the contrast ratio) is 25.371 $\pm 0.041$ (or 25.44 weighting the tip by the number of stars below the tip).  \citep{Jang22} finds 25.347 $\pm$ 0.015 (after MW extinction removal) from the outer region of a field overlapping field 3 from L23, which would propagate to the $\Delta H_0$ of 1.6 stated above with CCHP. On the other hand, EDD finds 25.402 $\pm 0.025$ from a combination of different fields, 5 and 6 from L23, though EDD applies a color correction which lowers the luminosity of the tip by $\sim$ 0.02 mag so that the difference in calibration with this study and EDD is -0.3 km/s/Mpc.  Without standardization, differences seen in the tip around NGC 4258 may be subject to the vagaries of individual field properties.  Specifically selecting the highest contrast (i.e., brightest) fields in NGC 4258 without a similar selection in the SN fields would lead to a bias in $H_0$.

More challenging is the ambiguity related to EDR tip selection. As discussed in \S \ref{subsec:notcr}, there are several, reasonable ways to identify the tip for a lumpy EDR of distant galaxies; as we report in Table~\ref{tab:neglect}, these can change $H_0$ by up to 5 km/s/Mpc  (e.g., between straight averaging peaks vs. highest EDR peak) if the tips are not standardized and subjected to quality criteria to determine their individual precisions.  It is important to define an algorithm for doing this {\it a priori} to avoid the potential for bias that may result from a case-by-case treatment.  Absent an unsupervised algorithm, a low dispersion between SN Ia and the TRGB could result from utilizing additional information in the CMD or by selecting the best tip with some dependence of the knowledge of the SN brightness. As a result, such low dispersion without an algorithmic approach may be unrealistic, or at least not an independent measure of TRGB precision.

We note that we use directly the photometry from \cite{Anand22}, but we expect that the photometry from \cite{Freedman19b}, if available, would be similar \citep{Jang22}. 
Finally, we remark that a central finding for this analysis and the preceding one from W22 is that TRGB from edge detection must be treated as a standardizable, rather than standard, candle.  A recent study by \cite{Anderson23} reinforced this viewpoint by measuring at $5\sigma$ two sub-populations of RGB stars based on their variability.  Remeasuring the absolute brightness of TRGB from LMC, but keeping the rest of the CCHP analysis, they find a higher value of $H_0$ of $71.8\pm1.5$ km/s/Mpc which is similar to our finding after excluding SN-related differences.  These works indicate that the TRGB measured from EDR should not be assumed to be a standard candle (or unbiased by the measurement process) but rather can and should be standardized using empirical measures calibrated from a large set of fields around the same host.

\section{Conclusions}

This analysis applies a standardized TRGB relation over a narrow color range to constrain $H_0$.  We find a value of $H_0=$ \ghostH~for our nominal analysis, with a standard deviation of the variants of the analysis of $\sim1$ ~km/s/Mpc.  We quantify that the application of the TCR relation increases the inferred value of $H_0$ by $\sim1.4$ km/s/Mpc compared to previous studies, while also providing a robust methodology for dealing with multiple tip detections for fields.  We find that a larger difference compared to previous TRGB studies is due to the supernova component of the analysis: using the recent Pantheon$+$ dataset increases $H_0$ by $\sim2.0$ km/s/Mpc. Our results address arguably the most pressing challenge to the validity of the Hubble tension, in that some previous TRGB results appeared to straddle the range between late-universe and early-universe measurements of $ H_0 $.  This result clarifies this tension further as one of the most pressing problems in cosmology today.

\begin{acknowledgments}

DS is supported by Department of Energy grant DE-SC0010007, the David and Lucile Packard Foundation, the Templeton Foundation and Sloan Foundation.  We greatly appreciate the GHOSTS team and EDD team for making all of their data public. S.D. acknowledges support from the Marie Curie Individual Fellowship under grant ID 890695 and a Junior Research Fellowship at Lucy Cavendish College. RIA is funded by the SNSF through an Eccellenza Professorial Fellowship, grant number PCEFP2\_194638.
RLB acknowledges support from NSF-AST 2108616. 

 This research has made use of NASA’s Astrophysics Data System.

\end{acknowledgments}

\section{Data Availability}

 We include all of the data and code used for this analysis.  Photometry for the SN hosts and NGC 4258 is based off the EDD CMDs/TRGB catalog \citep{Tully2009,Anand21b} pipeline and is accessible at \url{https://edd.ifa.hawaii.edu}. The training data comes from the GHOSTS program \citep{Radburn_Smith_2011} and is accessible at \url{https://archive.stsci.edu/prepds/ghosts/ghosts/survey.html}.  For ease, we also re-release this data, along with our codes at \url{https://github.com/JiaxiWu1018/CATS-H0}.

\facilities{MAST, HST:ACS}

\software{Astropy \citep{2013A&A...558A..33A, 2018AJ....156..123A},
          Matplotlib \citep{2007CSE.....9...90H},
          NumPy \citep{numpy, 2020Natur.585..357H},
}
\newpage

\begin{deluxetable*}{lcc}[t]
\tabletypesize{\small}
\tablecaption{Sources of Differences in H$_0$ Between TRGB analysis by CATs (Here), CCHP, EDD  (in $H_0$)\label{tb:trgbdiff}}
\tablehead{\colhead{Term} & \multicolumn{1}{c}{\hspace{0.5in}$\Delta$CCHP\ \ \ \ \ \ \ \ \ \ } & \multicolumn{1}{c}{\hspace{0.5in}$\Delta$EDD\ \ \ \ \ \ \ \ \ \ }} 
\startdata
\multicolumn{1}{c}{ }&{(km/s/Mpc)}&{(km/s/Mpc)} \\
\hline
\multicolumn{3}{c}{SN Related} \\
\hline
1. Include SN 2021pit,2021rhu,2007on  & 0.6 & 1.3 \\
2. No TRGB detected in N5584,N3021,N1309,N3370 \hspace{2in} & 0.0 & 0.0 \\
3. Peculiar Flows (Pantheon$+$) & 0.4 & 0.0 \\
4. Hubble Flow Surveys (Pantheon$+$) & 1.1 & 0.0 \\
\hline
\textbf{SN subtotal} & \textbf{2.0} & \textbf{1.3}  \\
\hline
\multicolumn{3}{c}{TRGB Related} \\
\hline
5. Fiducial TRGB Calibration/Tip-Contrast Relation & $1.4$ & $-0.3$  \\
\hline
\textbf{Total} & \textbf{3.4} & \textbf{1.0} \\
\enddata
\tablecomments{$\Delta$CCHP= differences between\citet{Freedman21} and EDD= differences with \citet{Anand22}, respectively.  Descriptions of individual entries: (1) CCHP did not include the two SNe from 2021 and excluded SN 2007on, EDD didn't include the two from 2021 (2) Neither CATs nor EDD detected the TRGB in these four most-distant SN host galaxies; (3) Pantheon$+$ accounts for peculiar motions which produce a highly significant improvement in the dispersion of the Hubble diagram (see \citealp{Peterson22})  (4) CCHP measures the Hubble flow from a single SN survey which has an offset with respect to the mean of many surveys in Pantheon$+$ as shown in \cite{Brownsberger22}; (5) This term is difference in the calibration of TRGB from NGC 4258 as applied to SN hosts and is discussed in \S 4.}
\label{tab:h0differences}
\end{deluxetable*}

\bibliography{paper}

\begin{thebibliography}{}
\expandafter\ifx\csname natexlab\endcsname\relax\def\natexlab#1{#1}\fi
\providecommand{\url}[1]{\href{#1}{#1}}
\providecommand{\dodoi}[1]{doi:~\href{http://doi.org/#1}{\nolinkurl{#1}}}
\providecommand{\doeprint}[1]{\href{http://ascl.net/#1}{\nolinkurl{http://ascl.net/#1}}}
\providecommand{\doarXiv}[1]{\href{https://arxiv.org/abs/#1}{\nolinkurl{https://arxiv.org/abs/#1}}}

\bibitem[{{Ahumada} {et~al.}(2020){Ahumada}, {Allende Prieto}, {Almeida},
  {Anders}, {Anderson}, {Andrews}, {Anguiano}, {Arcodia}, {Armengaud},
  {Aubert}, {Avila}, {Avila-Reese}, {Badenes}, {Balland}, {Barger},
  {Barrera-Ballesteros}, {Basu}, {Bautista}, {Beaton}, {Beers}, {Benavides},
  {Bender}, {Bernardi}, {Bershady}, {Beutler}, {Bidin}, {Bird}, {Bizyaev},
  {Blanc}, {Blanton}, {Boquien}, {Borissova}, {Bovy}, {Brandt}, {Brinkmann},
  {Brownstein}, {Bundy}, {Bureau}, {Burgasser}, {Burtin}, {Cano-D{\'\i}az},
  {Capasso}, {Cappellari}, {Carrera}, {Chabanier}, {Chaplin}, {Chapman},
  {Cherinka}, {Chiappini}, {Doohyun Choi}, {Chojnowski}, {Chung}, {Clerc},
  {Coffey}, {Comerford}, {Comparat}, {da Costa}, {Cousinou}, {Covey}, {Crane},
  {Cunha}, {Ilha}, {Dai}, {Damsted}, {Darling}, {Davidson}, {Davies}, {Dawson},
  {De}, {de la Macorra}, {De Lee}, {Queiroz}, {Deconto Machado}, {de la Torre},
  {Dell'Agli}, {du Mas des Bourboux}, {Diamond-Stanic}, {Dillon}, {Donor},
  {Drory}, {Duckworth}, {Dwelly}, {Ebelke}, {Eftekharzadeh}, {Davis Eigenbrot},
  {Elsworth}, {Eracleous}, {Erfanianfar}, {Escoffier}, {Fan}, {Farr},
  {Fern{\'a}ndez-Trincado}, {Feuillet}, {Finoguenov}, {Fofie},
  {Fraser-McKelvie}, {Frinchaboy}, {Fromenteau}, {Fu}, {Galbany}, {Garcia},
  {Garc{\'\i}a-Hern{\'a}ndez}, {Garma Oehmichen}, {Ge}, {Geimba Maia},
  {Geisler}, {Gelfand}, {Goddy}, {Gonzalez-Perez}, {Grabowski}, {Green},
  {Grier}, {Guo}, {Guy}, {Harding}, {Hasselquist}, {Hawken}, {Hayes}, {Hearty},
  {Hekker}, {Hogg}, {Holtzman}, {Horta}, {Hou}, {Hsieh}, {Huber}, {Hunt}, {Ider
  Chitham}, {Imig}, {Jaber}, {Jimenez Angel}, {Johnson}, {Jones},
  {J{\"o}nsson}, {Jullo}, {Kim}, {Kinemuchi}, {Kirkpatrick}, {Kite}, {Klaene},
  {Kneib}, {Kollmeier}, {Kong}, {Kounkel}, {Krishnarao}, {Lacerna}, {Lan},
  {Lane}, {Law}, {Le Goff}, {Leung}, {Lewis}, {Li}, {Lian}, {Lin}, {Long},
  {Longa-Pe{\~n}a}, {Lundgren}, {Lyke}, {Mackereth}, {MacLeod}, {Majewski},
  {Manchado}, {Maraston}, {Martini}, {Masseron}, {Masters}, {Mathur},
  {McDermid}, {Merloni}, {Merrifield}, {M{\'e}sz{\'a}ros}, {Miglio}, {Minniti},
  {Minsley}, {Miyaji}, {Mohammad}, {Mosser}, {Mueller}, {Muna},
  {Mu{\~n}oz-Guti{\'e}rrez}, {Myers}, {Nadathur}, {Nair}, {Nandra}, {Correa do
  Nascimento}, {Nevin}, {Newman}, {Nidever}, {Nitschelm}, {Noterdaeme},
  {O'Connell}, {Olmstead}, {Oravetz}, {Oravetz}, {Osorio}, {Pace}, {Padilla},
  {Palanque-Delabrouille}, {Palicio}, {Pan}, {Pan}, {Parker}, {Paviot},
  {Peirani}, {Ram{\'r}ez}, {Penny}, {Percival}, {Perez-Fournon},
  {P{\'e}rez-R{\`a}fols}, {Petitjean}, {Pieri}, {Pinsonneault}, {Poovelil},
  {Povick}, {Prakash}, {Price-Whelan}, {Raddick}, {Raichoor}, {Ray}, {Rembold},
  {Rezaie}, {Riffel}, {Riffel}, {Rix}, {Robin}, {Roman-Lopes},
  {Rom{\'a}n-Z{\'u}{\~n}iga}, {Rose}, {Ross}, {Rossi}, {Rowlands}, {Rubin},
  {Salvato}, {S{\'a}nchez}, {S{\'a}nchez-Menguiano}, {S{\'a}nchez-Gallego},
  {Sayres}, {Schaefer}, {Schiavon}, {Schimoia}, {Schlafly}, {Schlegel},
  {Schneider}, {Schultheis}, {Schwope}, {Seo}, {Serenelli}, {Shafieloo},
  {Shamsi}, {Shao}, {Shen}, {Shetrone}, {Shirley}, {Silva Aguirre}, {Simon},
  {Skrutskie}, {Slosar}, {Smethurst}, {Sobeck}, {Sodi}, {Souto}, {Stark},
  {Stassun}, {Steinmetz}, {Stello}, {Stermer}, {Storchi-Bergmann},
  {Streblyanska}, {Stringfellow}, {Stutz}, {Su{\'a}rez}, {Sun},
  {Taghizadeh-Popp}, {Talbot}, {Tayar}, {Thakar}, {Theriault}, {Thomas},
  {Thomas}, {Tinker}, {Tojeiro}, {Toledo}, {Tremonti}, {Troup}, {Tuttle},
  {Unda-Sanzana}, {Valentini}, {Vargas-Gonz{\'a}lez}, {Vargas-Maga{\~n}a},
  {V{\'a}zquez-Mata}, {Vivek}, {Wake}, {Wang}, {Weaver}, {Weijmans}, {Wild},
  {Wilson}, {Wilson}, {Wolthuis}, {Wood-Vasey}, {Yan}, {Yang}, {Y{\`e}che},
  {Zamora}, {Zarrouk}, {Zasowski}, {Zhang}, {Zhao}, {Zhao}, {Zheng}, {Zheng},
  {Zhu}, \& {Zou}}]{SDSS20}
{Ahumada}, R., {Allende Prieto}, C., {Almeida}, A., {et~al.} 2020, \apjs, 249,
  3, \dodoi{10.3847/1538-4365/ab929e}

\bibitem[{{Anand} {et~al.}(2022){Anand}, {Tully}, {Rizzi}, {Riess}, \&
  {Yuan}}]{Anand22}
{Anand}, G.~S., {Tully}, R.~B., {Rizzi}, L., {Riess}, A.~G., \& {Yuan}, W.
  2022, \apj, 932, 15, \dodoi{10.3847/1538-4357/ac68df}

\bibitem[{{Anand} {et~al.}(2021{\natexlab{a}}){Anand}, {Rizzi}, {Tully},
  {Shaya}, {Karachentsev}, {Makarov}, {Makarova}, {Wu}, {Dolphin}, \&
  {Kourkchi}}]{Anand21b}
{Anand}, G.~S., {Rizzi}, L., {Tully}, R.~B., {et~al.} 2021{\natexlab{a}}, \aj,
  162, 80, \dodoi{10.3847/1538-3881/ac0440}

\bibitem[{{Anand} {et~al.}(2021{\natexlab{b}}){Anand}, {Lee}, {Van Dyk},
  {Leroy}, {Rosolowsky}, {Schinnerer}, {Larson}, {Kourkchi}, {Kreckel},
  {Scheuermann}, {Rizzi}, {Thilker}, {Tully}, {Bigiel}, {Blanc}, {Boquien},
  {Chandar}, {Dale}, {Emsellem}, {Deger}, {Glover}, {Grasha}, {Groves}, {S.
  Klessen}, {Kruijssen}, {Querejeta}, {S{\'a}nchez-Bl{\'a}zquez}, {Schruba},
  {Turner}, {Ubeda}, {Williams}, \& {Whitmore}}]{Anand21}
{Anand}, G.~S., {Lee}, J.~C., {Van Dyk}, S.~D., {et~al.} 2021{\natexlab{b}},
  \mnras, 501, 3621, \dodoi{10.1093/mnras/staa3668}

\bibitem[{{Anderson}(2022)}]{Anderson2022}
{Anderson}, R.~I. 2022, \aap, 658, A148, \dodoi{10.1051/0004-6361/202141644}

\bibitem[{{Anderson} {et~al.}(2023){Anderson}, {Koblischke}, \&
  {Eyer}}]{Anderson23}
{Anderson}, R.~I., {Koblischke}, N.~W., \& {Eyer}, L. 2023, arXiv e-prints,
  arXiv:2303.04790.
\newblock \doarXiv{2303.04790}

\bibitem[{{Astropy Collaboration} {et~al.}(2013){Astropy Collaboration},
  {Robitaille}, {Tollerud}, {Greenfield}, {Droettboom}, {Bray}, {Aldcroft},
  {Davis}, {Ginsburg}, {Price-Whelan}, {Kerzendorf}, {Conley}, {Crighton},
  {Barbary}, {Muna}, {Ferguson}, {Grollier}, {Parikh}, {Nair}, {Unther},
  {Deil}, {Woillez}, {Conseil}, {Kramer}, {Turner}, {Singer}, {Fox}, {Weaver},
  {Zabalza}, {Edwards}, {Azalee Bostroem}, {Burke}, {Casey}, {Crawford},
  {Dencheva}, {Ely}, {Jenness}, {Labrie}, {Lim}, {Pierfederici}, {Pontzen},
  {Ptak}, {Refsdal}, {Servillat}, \& {Streicher}}]{2013A&A...558A..33A}
{Astropy Collaboration}, {Robitaille}, T.~P., {Tollerud}, E.~J., {et~al.} 2013,
  \aap, 558, A33, \dodoi{10.1051/0004-6361/201322068}

\bibitem[{{Astropy Collaboration} {et~al.}(2018){Astropy Collaboration},
  {Price-Whelan}, {Sip{\H{o}}cz}, {G{\"u}nther}, {Lim}, {Crawford}, {Conseil},
  {Shupe}, {Craig}, {Dencheva}, {Ginsburg}, {VanderPlas}, {Bradley},
  {P{\'e}rez-Su{\'a}rez}, {de Val-Borro}, {Aldcroft}, {Cruz}, {Robitaille},
  {Tollerud}, {Ardelean}, {Babej}, {Bach}, {Bachetti}, {Bakanov}, {Bamford},
  {Barentsen}, {Barmby}, {Baumbach}, {Berry}, {Biscani}, {Boquien}, {Bostroem},
  {Bouma}, {Brammer}, {Bray}, {Breytenbach}, {Buddelmeijer}, {Burke},
  {Calderone}, {Cano Rodr{\'\i}guez}, {Cara}, {Cardoso}, {Cheedella}, {Copin},
  {Corrales}, {Crichton}, {D'Avella}, {Deil}, {Depagne}, {Dietrich}, {Donath},
  {Droettboom}, {Earl}, {Erben}, {Fabbro}, {Ferreira}, {Finethy}, {Fox},
  {Garrison}, {Gibbons}, {Goldstein}, {Gommers}, {Greco}, {Greenfield},
  {Groener}, {Grollier}, {Hagen}, {Hirst}, {Homeier}, {Horton}, {Hosseinzadeh},
  {Hu}, {Hunkeler}, {Ivezi{\'c}}, {Jain}, {Jenness}, {Kanarek}, {Kendrew},
  {Kern}, {Kerzendorf}, {Khvalko}, {King}, {Kirkby}, {Kulkarni}, {Kumar},
  {Lee}, {Lenz}, {Littlefair}, {Ma}, {Macleod}, {Mastropietro}, {McCully},
  {Montagnac}, {Morris}, {Mueller}, {Mumford}, {Muna}, {Murphy}, {Nelson},
  {Nguyen}, {Ninan}, {N{\"o}the}, {Ogaz}, {Oh}, {Parejko}, {Parley}, {Pascual},
  {Patil}, {Patil}, {Plunkett}, {Prochaska}, {Rastogi}, {Reddy Janga},
  {Sabater}, {Sakurikar}, {Seifert}, {Sherbert}, {Sherwood-Taylor}, {Shih},
  {Sick}, {Silbiger}, {Singanamalla}, {Singer}, {Sladen}, {Sooley},
  {Sornarajah}, {Streicher}, {Teuben}, {Thomas}, {Tremblay}, {Turner},
  {Terr{\'o}n}, {van Kerkwijk}, {de la Vega}, {Watkins}, {Weaver}, {Whitmore},
  {Woillez}, {Zabalza}, \& {Astropy Contributors}}]{2018AJ....156..123A}
{Astropy Collaboration}, {Price-Whelan}, A.~M., {Sip{\H{o}}cz}, B.~M., {et~al.}
  2018, \aj, 156, 123, \dodoi{10.3847/1538-3881/aabc4f}

\bibitem[{{Beaton} {et~al.}(2018){Beaton}, {Bono}, {Braga}, {Dall'Ora},
  {Fiorentino}, {Jang}, {Mart{\'\i}nez-V{\'a}zquez}, {Matsunaga}, {Monelli},
  {Neeley}, \& {Salaris}}]{2018SSRv..214..113B}
{Beaton}, R.~L., {Bono}, G., {Braga}, V.~F., {et~al.} 2018, \ssr, 214, 113,
  \dodoi{10.1007/s11214-018-0542-1}

\bibitem[{{Betoule} {et~al.}(2014){Betoule}, {Kessler}, {Guy}, {Mosher},
  {Hardin}, {Biswas}, {Astier}, {El-Hage}, {Konig}, {Kuhlmann}, {Marriner},
  {Pain}, {Regnault}, {Balland}, {Bassett}, {Brown}, {Campbell}, {Carlberg},
  {Cellier-Holzem}, {Cinabro}, {Conley}, {D'Andrea}, {DePoy}, {Doi}, {Ellis},
  {Fabbro}, {Filippenko}, {Foley}, {Frieman}, {Fouchez}, {Galbany}, {Goobar},
  {Gupta}, {Hill}, {Hlozek}, {Hogan}, {Hook}, {Howell}, {Jha}, {Le Guillou},
  {Leloudas}, {Lidman}, {Marshall}, {M{\"o}ller}, {Mour{\~a}o}, {Neveu},
  {Nichol}, {Olmstead}, {Palanque-Delabrouille}, {Perlmutter}, {Prieto},
  {Pritchet}, {Richmond}, {Riess}, {Ruhlmann-Kleider}, {Sako}, {Schahmaneche},
  {Schneider}, {Smith}, {Sollerman}, {Sullivan}, {Walton}, \&
  {Wheeler}}]{Betoule14}
{Betoule}, M., {Kessler}, R., {Guy}, J., {et~al.} 2014, \aap, 568, A22,
  \dodoi{10.1051/0004-6361/201423413}

\bibitem[{{Blakeslee} {et~al.}(2021){Blakeslee}, {Jensen}, {Ma}, {Milne}, \&
  {Greene}}]{Blakeslee21}
{Blakeslee}, J.~P., {Jensen}, J.~B., {Ma}, C.-P., {Milne}, P.~A., \& {Greene},
  J.~E. 2021, \apj, 911, 65, \dodoi{10.3847/1538-4357/abe86a}

\bibitem[{{Brout} {et~al.}(2022{\natexlab{a}}){Brout}, {Scolnic}, {Popovic},
  {Riess}, {Carr}, {Zuntz}, {Kessler}, {Davis}, {Hinton}, {Jones}, {Kenworthy},
  {Peterson}, {Said}, {Taylor}, {Ali}, {Armstrong}, {Charvu}, {Dwomoh},
  {Meldorf}, {Palmese}, {Qu}, {Rose}, {Sanchez}, {Stubbs}, {Vincenzi}, {Wood},
  {Brown}, {Chen}, {Chambers}, {Coulter}, {Dai}, {Dimitriadis}, {Filippenko},
  {Foley}, {Jha}, {Kelsey}, {Kirshner}, {M{\"o}ller}, {Muir}, {Nadathur},
  {Pan}, {Rest}, {Rojas-Bravo}, {Sako}, {Siebert}, {Smith}, {Stahl}, \&
  {Wiseman}}]{Brout22a}
{Brout}, D., {Scolnic}, D., {Popovic}, B., {et~al.} 2022{\natexlab{a}}, \apj,
  938, 110, \dodoi{10.3847/1538-4357/ac8e04}

\bibitem[{{Brout} {et~al.}(2022{\natexlab{b}}){Brout}, {Taylor}, {Scolnic},
  {Wood}, {Rose}, {Vincenzi}, {Dwomoh}, {Lidman}, {Riess}, {Ali}, {Qu}, \&
  {Dai}}]{Brout22b}
{Brout}, D., {Taylor}, G., {Scolnic}, D., {et~al.} 2022{\natexlab{b}}, \apj,
  938, 111, \dodoi{10.3847/1538-4357/ac8bcc}

\bibitem[{{Brout} {et~al.}(2022{\natexlab{c}}){Brout}, {Scolnic}, {Popovic},
  {Riess}, {Zuntz}, {Kessler}, {Carr}, {Davis}, {Hinton}, {Jones}, {Kenworthy},
  {Peterson}, {Said}, {Taylor}, {Ali}, {Armstrong}, {Charvu}, {Dwomoh},
  {Palmese}, {Qu}, {Rose}, {Stubbs}, {Vincenzi}, {Wood}, {Brown}, {Chen},
  {Chambers}, {Coulter}, {Dai}, {Dimitriadis}, {Filippenko}, {Foley}, {Jha},
  {Kelsey}, {Kirshner}, {M{\"o}ller}, {Muir}, {Nadathur}, {Pan}, {Rest},
  {Rojas-Bravo}, {Sako}, {Siebert}, {Smith}, {Stahl}, \& {Wiseman}}]{Brout22}
{Brout}, D., {Scolnic}, D., {Popovic}, B., {et~al.} 2022{\natexlab{c}}, arXiv
  e-prints, arXiv:2202.04077.
\newblock \doarXiv{2202.04077}

\bibitem[{{Brown} {et~al.}(2014){Brown}, {Breeveld}, {Holland}, {Kuin}, \&
  {Pritchard}}]{Swift14}
{Brown}, P.~J., {Breeveld}, A.~A., {Holland}, S., {Kuin}, P., \& {Pritchard},
  T. 2014, \apss, 354, 89, \dodoi{10.1007/s10509-014-2059-8}

\bibitem[{{Brownsberger} {et~al.}(2021){Brownsberger}, {Brout}, {Scolnic},
  {Stubbs}, \& {Riess}}]{Brownsberger22}
{Brownsberger}, S., {Brout}, D., {Scolnic}, D., {Stubbs}, C.~W., \& {Riess},
  A.~G. 2021, arXiv e-prints, arXiv:2110.03486.
\newblock \doarXiv{2110.03486}

\bibitem[{{Capozzi} \& {Raffelt}(2020)}]{Capozzi20}
{Capozzi}, F., \& {Raffelt}, G. 2020, \prd, 102, 083007,
  \dodoi{10.1103/PhysRevD.102.083007}

\bibitem[{{Carr} {et~al.}(2022){Carr}, {Davis}, {Scolnic}, {Said}, {Brout},
  {Peterson}, \& {Kessler}}]{Carr22}
{Carr}, A., {Davis}, T.~M., {Scolnic}, D., {et~al.} 2022, \pasa, 39, e046,
  \dodoi{10.1017/pasa.2022.41}

\bibitem[{{Choi} {et~al.}(2016){Choi}, {Dotter}, {Conroy}, {Cantiello},
  {Paxton}, \& {Johnson}}]{mist16}
{Choi}, J., {Dotter}, A., {Conroy}, C., {et~al.} 2016, \apj, 823, 102,
  \dodoi{10.3847/0004-637X/823/2/102}

\bibitem[{{Dhawan} {et~al.}(2022){Dhawan}, {Goobar}, {Johansson}, {Jang},
  {Rigault}, {Harvey}, {Maguire}, {Freedman}, {Madore}, {Smith}, {Sollerman},
  {Kim}, {Andreoni}, {Bellm}, {Coughlin}, {Dekany}, {Graham}, {Kulkarni},
  {Laher}, {Medford}, {Neill}, {Nir}, {Riddle}, \& {Rusholme}}]{Dhawan22}
{Dhawan}, S., {Goobar}, A., {Johansson}, J., {et~al.} 2022, \apj, 934, 185,
  \dodoi{10.3847/1538-4357/ac7ceb}

\bibitem[{{Dolphin}(2016)}]{DOLPHOT}
{Dolphin}, A. 2016, {DOLPHOT: Stellar photometry}, Astrophysics Source Code
  Library, record ascl:1608.013.
\newblock \doeprint{1608.013}

\bibitem[{{Freedman}(2021)}]{Freedman21}
{Freedman}, W.~L. 2021, \apj, 919, 16, \dodoi{10.3847/1538-4357/ac0e95}

\bibitem[{Freedman {et~al.}(2019)Freedman, Madore, Hatt, Hoyt, Jang, Beaton,
  Burns, Lee, Monson, Neeley, Phillips, Rich, \& Seibert}]{Freedman19b}
Freedman, W.~L., Madore, B.~F., Hatt, D., {et~al.} 2019, The Astrophysical
  Journal, 882, 34, \dodoi{10.3847/1538-4357/ab2f73}

\bibitem[{{Freedman} {et~al.}(2020){Freedman}, {Madore}, {Hoyt}, {Jang},
  {Beaton}, {Lee}, {Monson}, {Neeley}, \& {Rich}}]{Freedman20}
{Freedman}, W.~L., {Madore}, B.~F., {Hoyt}, T., {et~al.} 2020, \apj, 891, 57,
  \dodoi{10.3847/1538-4357/ab7339}

\bibitem[{{Harris} {et~al.}(2020){Harris}, {Millman}, {van der Walt},
  {Gommers}, {Virtanen}, {Cournapeau}, {Wieser}, {Taylor}, {Berg}, {Smith},
  {Kern}, {Picus}, {Hoyer}, {van Kerkwijk}, {Brett}, {Haldane}, {del R{\'\i}o},
  {Wiebe}, {Peterson}, {G{\'e}rard-Marchant}, {Sheppard}, {Reddy}, {Weckesser},
  {Abbasi}, {Gohlke}, \& {Oliphant}}]{2020Natur.585..357H}
{Harris}, C.~R., {Millman}, K.~J., {van der Walt}, S.~J., {et~al.} 2020, \nat,
  585, 357, \dodoi{10.1038/s41586-020-2649-2}

\bibitem[{Hatt {et~al.}(2017)Hatt, Beaton, Freedman, Madore, Jang, Hoyt, Lee,
  Monson, Rich, Scowcroft, \& Seibert}]{Hatt_2017}
Hatt, D., Beaton, R.~L., Freedman, W.~L., {et~al.} 2017, The Astrophysical
  Journal, 845, 146, \dodoi{10.3847/1538-4357/aa7f73}

\bibitem[{{Hoyt}(2023)}]{Hoyt23}
{Hoyt}, T.~J. 2023, Nature Astronomy, \dodoi{10.1038/s41550-023-01913-1}

\bibitem[{Hoyt {et~al.}(2021)Hoyt, Beaton, Freedman, Jang, Lee, Madore, Monson,
  Neeley, Rich, \& Seibert}]{Hoyt_2021}
Hoyt, T.~J., Beaton, R.~L., Freedman, W.~L., {et~al.} 2021, The Astrophysical
  Journal, 915, 34, \dodoi{10.3847/1538-4357/abfe5a}

\bibitem[{{Hunter}(2007)}]{2007CSE.....9...90H}
{Hunter}, J.~D. 2007, Computing in Science and Engineering, 9, 90,
  \dodoi{10.1109/MCSE.2007.55}

\bibitem[{{Jang}(2022)}]{Jang22}
{Jang}, I.~S. 2022, arXiv e-prints, arXiv:2208.02824.
\newblock \doarXiv{2208.02824}

\bibitem[{{Jang} \& {Lee}(2017)}]{Jang17}
{Jang}, I.~S., \& {Lee}, M.~G. 2017, \apj, 836, 74,
  \dodoi{10.3847/1538-4357/836/1/74}

\bibitem[{{Jang} {et~al.}(2018){Jang}, {Hatt}, {Beaton}, {Lee}, {Freedman},
  {Madore}, {Hoyt}, {Monson}, {Rich}, {Scowcroft}, \& {Seibert}}]{Jang_2018}
{Jang}, I.~S., {Hatt}, D., {Beaton}, R.~L., {et~al.} 2018, \apj, 852, 60,
  \dodoi{10.3847/1538-4357/aa9d92}

\bibitem[{{Jang} {et~al.}(2021){Jang}, {Hoyt}, {Beaton}, {Freedman}, {Madore},
  {Lee}, {Neeley}, {Monson}, {Rich}, \& {Seibert}}]{Jang21}
{Jang}, I.~S., {Hoyt}, T.~J., {Beaton}, R.~L., {et~al.} 2021, \apj, 906, 125,
  \dodoi{10.3847/1538-4357/abc8e9}

\bibitem[{{Jones} {et~al.}(2022){Jones}, {Mandel}, {Kirshner}, {Thorp},
  {Challis}, {Avelino}, {Brout}, {Burns}, {Foley}, {Pan}, {Scolnic}, {Siebert},
  {Chornock}, {Freedman}, {Friedman}, {Frieman}, {Galbany}, {Hsiao}, {Kelsey},
  {Marion}, {Nichol}, {Nugent}, {Phillips}, {Rest}, {Riess}, {Sako}, {Smith},
  {Wiseman}, \& {Wood-Vasey}}]{Jones22}
{Jones}, D.~O., {Mandel}, K.~S., {Kirshner}, R.~P., {et~al.} 2022, \apj, 933,
  172, \dodoi{10.3847/1538-4357/ac755b}

\bibitem[{{Kourkchi} {et~al.}(2022){Kourkchi}, {Tully}, {Courtois}, {Dupuy}, \&
  {Guinet}}]{Kourkchi22}
{Kourkchi}, E., {Tully}, R.~B., {Courtois}, H.~M., {Dupuy}, A., \& {Guinet}, D.
  2022, \mnras, 511, 6160, \dodoi{10.1093/mnras/stac303}

\bibitem[{{Krisciunas} {et~al.}(2017){Krisciunas}, {Contreras}, {Burns},
  {Phillips}, {Stritzinger}, {Morrell}, {Hamuy}, {Anais}, {Boldt}, {Busta},
  {Campillay}, {Castell{\'o}n}, {Folatelli}, {Freedman}, {Gonz{\'a}lez},
  {Hsiao}, {Krzeminski}, {Persson}, {Roth}, {Salgado}, {Ser{\'o}n}, {Suntzeff},
  {Torres}, {Filippenko}, {Li}, {Madore}, {DePoy}, {Marshall}, {Rheault}, \&
  {Villanueva}}]{krisciunas17}
{Krisciunas}, K., {Contreras}, C., {Burns}, C.~R., {et~al.} 2017, \aj, 154,
  211, \dodoi{10.3847/1538-3881/aa8df0}

\bibitem[{{Lee} {et~al.}(1993){Lee}, {Freedman}, \& {Madore}}]{Lee93}
{Lee}, M.~G., {Freedman}, W.~L., \& {Madore}, B.~F. 1993, \apj, 417, 553,
  \dodoi{10.1086/173334}

\bibitem[{{Li} {et~al.}(2022){Li}, {Casertano}, \& {Riess}}]{Li22}
{Li}, S., {Casertano}, S., \& {Riess}, A.~G. 2022, arXiv e-prints,
  arXiv:2202.11110.
\newblock \doarXiv{2202.11110}

\bibitem[{{Makarov} {et~al.}(2006){Makarov}, {Makarova}, {Rizzi}, {Tully},
  {Dolphin}, {Sakai}, \& {Shaya}}]{Makorov06}
{Makarov}, D., {Makarova}, L., {Rizzi}, L., {et~al.} 2006, \aj, 132, 2729,
  \dodoi{10.1086/508925}

\bibitem[{{McQuinn} {et~al.}(2021){McQuinn}, {Telidevara}, {Fuson}, {Adams},
  {Cannon}, {Skillman}, {Dolphin}, {Haynes}, {Rhode}, {Salzer}, {Giovanelli},
  \& {Gordon}}]{2021ApJ...918...23M}
{McQuinn}, K. B.~W., {Telidevara}, A.~K., {Fuson}, J., {et~al.} 2021, \apj,
  918, 23, \dodoi{10.3847/1538-4357/ac03ae}

\bibitem[{{M{\'e}nard} {et~al.}(2010){M{\'e}nard}, {Scranton}, {Fukugita}, \&
  {Richards}}]{Menard10}
{M{\'e}nard}, B., {Scranton}, R., {Fukugita}, M., \& {Richards}, G. 2010,
  \mnras, 405, 1025, \dodoi{10.1111/j.1365-2966.2010.16486.x}

\bibitem[{{M{\'e}ndez} {et~al.}(2002){M{\'e}ndez}, {Davis}, {Moustakas},
  {Newman}, {Madore}, \& {Freedman}}]{Mendez_2002}
{M{\'e}ndez}, B., {Davis}, M., {Moustakas}, J., {et~al.} 2002, \aj, 124, 213,
  \dodoi{10.1086/341168}

\bibitem[{{Mutlu-Pakdil} {et~al.}(2022){Mutlu-Pakdil}, {Sand}, {Crnojevi{\'c}},
  {Jones}, {Caldwell}, {Guhathakurta}, {Seth}, {Simon}, {Spekkens}, {Strader},
  \& {Toloba}}]{2022ApJ...926...77M}
{Mutlu-Pakdil}, B., {Sand}, D.~J., {Crnojevi{\'c}}, D., {et~al.} 2022, \apj,
  926, 77, \dodoi{10.3847/1538-4357/ac4418}

\bibitem[{{Pesce} {et~al.}(2020){Pesce}, {Braatz}, {Reid}, {Riess}, {Scolnic},
  {Condon}, {Gao}, {Henkel}, {Impellizzeri}, {Kuo}, \& {Lo}}]{Pesce20}
{Pesce}, D.~W., {Braatz}, J.~A., {Reid}, M.~J., {et~al.} 2020, \apjl, 891, L1,
  \dodoi{10.3847/2041-8213/ab75f0}

\bibitem[{{Peterson} {et~al.}(2022){Peterson}, {Kenworthy}, {Scolnic}, {Riess},
  {Brout}, {Carr}, {Courtois}, {Davis}, {Dwomoh}, {Jones}, {Popovic}, {Rose},
  \& {Said}}]{Peterson22}
{Peterson}, E.~R., {Kenworthy}, W.~D., {Scolnic}, D., {et~al.} 2022, \apj, 938,
  112, \dodoi{10.3847/1538-4357/ac4698}

\bibitem[{{Planck Collaboration} {et~al.}(2020){Planck Collaboration},
  {Aghanim}, {Akrami}, {Ashdown}, {Aumont}, {Baccigalupi}, {Ballardini},
  {Banday}, {Barreiro}, {Bartolo}, {Basak}, {Battye}, {Benabed}, {Bernard},
  {Bersanelli}, {Bielewicz}, {Bock}, {Bond}, {Borrill}, {Bouchet}, {Boulanger},
  {Bucher}, {Burigana}, {Butler}, {Calabrese}, {Cardoso}, {Carron},
  {Challinor}, {Chiang}, {Chluba}, {Colombo}, {Combet}, {Contreras}, {Crill},
  {Cuttaia}, {de Bernardis}, {de Zotti}, {Delabrouille}, {Delouis}, {Di
  Valentino}, {Diego}, {Dor{\'e}}, {Douspis}, {Ducout}, {Dupac}, {Dusini},
  {Efstathiou}, {Elsner}, {En{\ss}lin}, {Eriksen}, {Fantaye}, {Farhang},
  {Fergusson}, {Fernandez-Cobos}, {Finelli}, {Forastieri}, {Frailis},
  {Fraisse}, {Franceschi}, {Frolov}, {Galeotta}, {Galli}, {Ganga},
  {G{\'e}nova-Santos}, {Gerbino}, {Ghosh}, {Gonz{\'a}lez-Nuevo}, {G{\'o}rski},
  {Gratton}, {Gruppuso}, {Gudmundsson}, {Hamann}, {Handley}, {Hansen},
  {Herranz}, {Hildebrandt}, {Hivon}, {Huang}, {Jaffe}, {Jones}, {Karakci},
  {Keih{\"a}nen}, {Keskitalo}, {Kiiveri}, {Kim}, {Kisner}, {Knox},
  {Krachmalnicoff}, {Kunz}, {Kurki-Suonio}, {Lagache}, {Lamarre}, {Lasenby},
  {Lattanzi}, {Lawrence}, {Le Jeune}, {Lemos}, {Lesgourgues}, {Levrier},
  {Lewis}, {Liguori}, {Lilje}, {Lilley}, {Lindholm}, {L{\'o}pez-Caniego},
  {Lubin}, {Ma}, {Mac{\'\i}as-P{\'e}rez}, {Maggio}, {Maino}, {Mandolesi},
  {Mangilli}, {Marcos-Caballero}, {Maris}, {Martin}, {Martinelli},
  {Mart{\'\i}nez-Gonz{\'a}lez}, {Matarrese}, {Mauri}, {McEwen}, {Meinhold},
  {Melchiorri}, {Mennella}, {Migliaccio}, {Millea}, {Mitra},
  {Miville-Desch{\^e}nes}, {Molinari}, {Montier}, {Morgante}, {Moss}, {Natoli},
  {N{\o}rgaard-Nielsen}, {Pagano}, {Paoletti}, {Partridge}, {Patanchon},
  {Peiris}, {Perrotta}, {Pettorino}, {Piacentini}, {Polastri}, {Polenta},
  {Puget}, {Rachen}, {Reinecke}, {Remazeilles}, {Renzi}, {Rocha}, {Rosset},
  {Roudier}, {Rubi{\~n}o-Mart{\'\i}n}, {Ruiz-Granados}, {Salvati}, {Sandri},
  {Savelainen}, {Scott}, {Shellard}, {Sirignano}, {Sirri}, {Spencer},
  {Sunyaev}, {Suur-Uski}, {Tauber}, {Tavagnacco}, {Tenti}, {Toffolatti},
  {Tomasi}, {Trombetti}, {Valenziano}, {Valiviita}, {Van Tent}, {Vibert},
  {Vielva}, {Villa}, {Vittorio}, {Wandelt}, {Wehus}, {White}, {White},
  {Zacchei}, \& {Zonca}}]{Planck18}
{Planck Collaboration}, {Aghanim}, N., {Akrami}, Y., {et~al.} 2020, \aap, 641,
  A6, \dodoi{10.1051/0004-6361/201833910}

\bibitem[{Radburn-Smith {et~al.}(2011)Radburn-Smith, de~Jong, Seth, Bailin,
  Bell, Brown, Bullock, Courteau, Dalcanton, Ferguson, Goudfrooij, Holfeltz,
  Holwerda, Purcell, Sick, Streich, Vlajic, \& Zucker}]{Radburn_Smith_2011}
Radburn-Smith, D.~J., de~Jong, R.~S., Seth, A.~C., {et~al.} 2011, The
  Astrophysical Journal Supplement Series, 195, 18,
  \dodoi{10.1088/0067-0049/195/2/18}

\bibitem[{{Riess} {et~al.}(2022){Riess}, {Yuan}, {Macri}, {Scolnic}, {Brout},
  {Casertano}, {Jones}, {Murakami}, {Anand}, {Breuval}, {Brink}, {Filippenko},
  {Hoffmann}, {Jha}, {D'arcy Kenworthy}, {Mackenty}, {Stahl}, \&
  {Zheng}}]{Riess22}
{Riess}, A.~G., {Yuan}, W., {Macri}, L.~M., {et~al.} 2022, \apjl, 934, L7,
  \dodoi{10.3847/2041-8213/ac5c5b}

\bibitem[{{Rose} {et~al.}(2022){Rose}, {Popovic}, {Scolnic}, \&
  {Brout}}]{Rose22}
{Rose}, B.~M., {Popovic}, B., {Scolnic}, D., \& {Brout}, D. 2022, \mnras, 516,
  4822, \dodoi{10.1093/mnras/stac2500}

\bibitem[{{Schlafly} \& {Finkbeiner}(2011)}]{Schlafly11}
{Schlafly}, E.~F., \& {Finkbeiner}, D.~P. 2011, \apj, 737, 103,
  \dodoi{10.1088/0004-637X/737/2/103}

\bibitem[{{Scolnic} {et~al.}(2015){Scolnic}, {Casertano}, {Riess}, {Rest},
  {Schlafly}, {Foley}, {Finkbeiner}, {Tang}, {Burgett}, {Chambers}, {Draper},
  {Flewelling}, {Hodapp}, {Huber}, {Kaiser}, {Kudritzki}, {Magnier},
  {Metcalfe}, \& {Stubbs}}]{Scolnic15}
{Scolnic}, D., {Casertano}, S., {Riess}, A., {et~al.} 2015, \apj, 815, 117,
  \dodoi{10.1088/0004-637X/815/2/117}

\bibitem[{{Scolnic} {et~al.}(2020){Scolnic}, {Smith}, {Massiah}, {Wiseman},
  {Brout}, {Kessler}, {Davis}, {Foley}, {Galbany}, {Hinton}, {Hounsell},
  {Kelsey}, {Lidman}, {Macaulay}, {Morgan}, {Nichol}, {M{\"o}ller}, {Popovic},
  {Sako}, {Sullivan}, {Thomas}, {Tucker}, {Abbott}, {Aguena}, {Allam}, {Annis},
  {Avila}, {Bechtol}, {Bertin}, {Brooks}, {Burke}, {Rosell}, {Carollo}, {Kind},
  {Carretero}, {Costanzi}, {da Costa}, {De Vicente}, {Desai}, {Diehl}, {Doel},
  {Drlica-Wagner}, {Eckert}, {Eifler}, {Everett}, {Flaugher}, {Fosalba},
  {Frieman}, {Garc{\'\i}a-Bellido}, {Gaztanaga}, {Gerdes}, {Glazebrook},
  {Gruen}, {Gruendl}, {Gschwend}, {Gutierrez}, {Hartley}, {Hollowood},
  {Honscheid}, {James}, {Kuehn}, {Kuropatkin}, {Lewis}, {Li}, {Lima}, {Maia},
  {Marshall}, {Menanteau}, {Miquel}, {Palmese}, {Paz-Chinch{\'o}n}, {Plazas},
  {Pursiainen}, {Sanchez}, {Scarpine}, {Schubnell}, {Serrano},
  {Sevilla-Noarbe}, {Sommer}, {Suchyta}, {Swanson}, {Tarle}, {Varga}, {Walker},
  {Wilkinson}, \& {DES Collaboration}}]{Scolnic20}
{Scolnic}, D., {Smith}, M., {Massiah}, A., {et~al.} 2020, \apjl, 896, L13,
  \dodoi{10.3847/2041-8213/ab8735}

\bibitem[{{Scolnic} {et~al.}(2022){Scolnic}, {Brout}, {Carr}, {Riess}, {Davis},
  {Dwomoh}, {Jones}, {Ali}, {Charvu}, {Chen}, {Peterson}, {Popovic}, {Rose},
  {Wood}, {Brown}, {Chambers}, {Coulter}, {Dettman}, {Dimitriadis},
  {Filippenko}, {Foley}, {Jha}, {Kilpatrick}, {Kirshner}, {Pan}, {Rest},
  {Rojas-Bravo}, {Siebert}, {Stahl}, \& {Zheng}}]{Scolnic22}
{Scolnic}, D., {Brout}, D., {Carr}, A., {et~al.} 2022, \apj, 938, 113,
  \dodoi{10.3847/1538-4357/ac8b7a}

\bibitem[{{Scolnic} {et~al.}(2018){Scolnic}, {Jones}, {Rest}, {Pan},
  {Chornock}, {Foley}, {Huber}, {Kessler}, {Narayan}, {Riess}, {Rodney},
  {Berger}, {Brout}, {Challis}, {Drout}, {Finkbeiner}, {Lunnan}, {Kirshner},
  {Sanders}, {Schlafly}, {Smartt}, {Stubbs}, {Tonry}, {Wood-Vasey}, {Foley},
  {Hand}, {Johnson}, {Burgett}, {Chambers}, {Draper}, {Hodapp}, {Kaiser},
  {Kudritzki}, {Magnier}, {Metcalfe}, {Bresolin}, {Gall}, {Kotak}, {McCrum}, \&
  {Smith}}]{Scolnic18}
{Scolnic}, D.~M., {Jones}, D.~O., {Rest}, A., {et~al.} 2018, \apj, 859, 101,
  \dodoi{10.3847/1538-4357/aab9bb}

\bibitem[{{Serenelli} {et~al.}(2017){Serenelli}, {Weiss}, {Cassisi}, {Salaris},
  \& {Pietrinferni}}]{Serenelli:2017}
{Serenelli}, A., {Weiss}, A., {Cassisi}, S., {Salaris}, M., \& {Pietrinferni},
  A. 2017, \aap, 606, A33, \dodoi{10.1051/0004-6361/201731004}

\bibitem[{{Shaya} {et~al.}(2022){Shaya}, {Tully}, {Pomar{\`e}de}, \&
  {Peel}}]{2022ApJ...927..168S}
{Shaya}, E.~J., {Tully}, R.~B., {Pomar{\`e}de}, D., \& {Peel}, A. 2022, \apj,
  927, 168, \dodoi{10.3847/1538-4357/ac4f66}

\bibitem[{{Shen} {et~al.}(2021){Shen}, {Danieli}, {van Dokkum}, {Abraham},
  {Brodie}, {Conroy}, {Dolphin}, {Romanowsky}, {Kruijssen}, \& {Dutta
  Chowdhury}}]{2021ApJ...914L..12S}
{Shen}, Z., {Danieli}, S., {van Dokkum}, P., {et~al.} 2021, \apjl, 914, L12,
  \dodoi{10.3847/2041-8213/ac0335}

\bibitem[{{Soltis} {et~al.}(2021){Soltis}, {Casertano}, \& {Riess}}]{Soltis21}
{Soltis}, J., {Casertano}, S., \& {Riess}, A.~G. 2021, \apjl, 908, L5,
  \dodoi{10.3847/2041-8213/abdbad}

\bibitem[{{Stetson}(1987)}]{Stetson87}
{Stetson}, P.~B. 1987, \pasp, 99, 191, \dodoi{10.1086/131977}

\bibitem[{Tully {et~al.}(2009)Tully, Rizzi, Shaya, Courtois, Makarov, \&
  Jacobs}]{Tully2009}
Tully, R.~B., Rizzi, L., Shaya, E.~J., {et~al.} 2009, The Astronomical Journal,
  138, 323, \dodoi{10.1088/0004-6256/138/2/323}

\bibitem[{{Tully} {et~al.}(2023){Tully}, {Kourkchi}, {Courtois}, {Anand},
  {Blakeslee}, {Brout}, {Jaeger}, {Dupuy}, {Guinet}, {Howlett}, {Jensen},
  {Pomar{\`e}de}, {Rizzi}, {Rubin}, {Said}, {Scolnic}, \&
  {Stahl}}]{2023ApJ...944...94T}
{Tully}, R.~B., {Kourkchi}, E., {Courtois}, H.~M., {et~al.} 2023, \apj, 944,
  94, \dodoi{10.3847/1538-4357/ac94d8}

\bibitem[{{van der Walt} {et~al.}(2011){van der Walt}, Colbert, \&
  Varoquaux}]{numpy}
{van der Walt}, S., Colbert, S.~C., \& Varoquaux, G. 2011, Computing in Science
  \& Engineering, 13, 22, \dodoi{10.1109/MCSE.2011.37}

\bibitem[{{Visser}(2004)}]{Visser04}
{Visser}, M. 2004, Classical and Quantum Gravity, 21, 2603,
  \dodoi{10.1088/0264-9381/21/11/006}

\bibitem[{{Wu} {et~al.}(2022){Wu}, {Scolnic}, {Riess}, {Anand}, {Beaton},
  {Casertano}, {Ke}, \& {Li}}]{Wu22}
{Wu}, J., {Scolnic}, D., {Riess}, A.~G., {et~al.} 2022, arXiv e-prints,
  arXiv:2211.06354.
\newblock \doarXiv{2211.06354}

\bibitem[{{Wu} {et~al.}(2014){Wu}, {Tully}, {Rizzi}, {Dolphin}, {Jacobs}, \&
  {Karachentsev}}]{Wu14}
{Wu}, P.-F., {Tully}, R.~B., {Rizzi}, L., {et~al.} 2014, \aj, 148, 7,
  \dodoi{10.1088/0004-6256/148/1/7}

\end{thebibliography}
\bibliographystyle{aasjournal}

\begin{appendix}
\section{Data Table for Impact of Variants}

We present here a full table of all the variants considered in this analysis in Table~\ref{tab:variants}.  As discussed in \S \ref{subsec:changingparam}, we have four separate variant parameters.  For each run, we report the values of the variant parameters (column 1 - 4).  We then report the inferred value of $H_0$ and its statistical uncertainty (column 6), the mean value of the brightness difference between the SN magnitudes and TRGB brightnesses (column 7), the mean value of the brightness difference between the TRGB brightnesses of NGC 4258 and the geometric distance measurement (column 8), the $\chi^2/N$ of the residuals of the SN-TRGB rung (column 9)), the dispersion of the residuals in the SN-TRGB rung (column 10), the number of TRGB tips of SN hosts measured (column 11), the number of SN hosts with a tip measured (column 13), and the number SNe in those hosts.

\startlongtable
\tabletypesize{\scriptsize}
\begin{deluxetable*}{ |c c c c | c c c | c c c c c|  } 
\label{tab:variants}
\tablecaption{Analysis variants and their impact on recovered parameters }
\tablehead{\colhead{$\textbf{\textrm{Sp.Clip~\%}}$} & \colhead{$\textbf{\textrm{Width}}$} & \colhead{$\textbf{\textrm{Smooth}}~\sigma$}   & \colhead{$\textbf{\textrm{MinTh~\%}}$} & \colhead{$H_0$}    &   \colhead{$\overline{\Delta S}$} &    \colhead{$m_{I,N4258,TRGB}^{R=4}$}   & \colhead{$\chi^2/N$}  & \colhead{$\sigma$} & \colhead{SNtips} & \colhead{SNhosts} & \colhead{SN}}  
\startdata                       
$10\%$ & 0.75 & 0.07 & 0.5 & $71.93\pm1.85$ & $15.19\pm0.04$ & $25.30\pm0.01$ & 9.07 & 0.20 & 23 & 13 & 18\\ 
$10\%$ & 0.75 & 0.07 & 0.6 & $72.48\pm1.88$ & $15.21\pm0.04$ & $25.33\pm0.01$ & 7.73 & 0.20 & 21 & 13 & 18\\ 
$10\%$ & 0.75 & 0.07 & 0.8 & $74.02\pm1.93$ & $15.22\pm0.04$ & $25.39\pm0.01$ & 2.48 & 0.20 & 18 & 13 & 18\\ 
$10\%$ & 0.75 & 0.10 & 0.5 & $72.74\pm1.97$ & $15.23\pm0.05$ & $25.37\pm0.01$ & 1.36 & 0.21 & 18 & 13 & 18\\ 
$10\%$ & 0.75 & 0.10 & 0.6 & $72.74\pm1.98$ & $15.25\pm0.05$ & $25.39\pm0.01$ & 2.29 & 0.21 & 17 & 13 & 18\\ 
$10\%$ & 0.75 & 0.10 & 0.8 & $72.76\pm1.98$ & $15.25\pm0.05$ & $25.39\pm0.01$ & 2.49 & 0.21 & 16 & 13 & 18\\ 
$10\%$ & 0.75 & 0.15 & 0.5 & $71.80\pm1.91$ & $15.25\pm0.04$ & $25.35\pm0.01$ & 1.51 & 0.18 & 13 & 13 & 18\\ 
$10\%$ & 0.75 & 0.15 & 0.6 & $71.81\pm1.91$ & $15.25\pm0.04$ & $25.35\pm0.01$ & 1.51 & 0.18 & 13 & 13 & 18\\ 
$10\%$ & 0.75 & 0.15 & 0.8 & $71.81\pm1.91$ & $15.25\pm0.04$ & $25.35\pm0.01$ & 1.51 & 0.18 & 13 & 13 & 18\\ 
$10\%$ & 1.0 & 0.07 & 0.5 & $71.74\pm1.93$ & $15.21\pm0.05$ & $25.32\pm0.01$ & 21.11 & 0.21 & 22 & 13 & 18\\ 
$10\%$ & 1.0 & 0.07 & 0.6 & $71.83\pm1.97$ & $15.23\pm0.05$ & $25.34\pm0.02$ & 25.27 & 0.21 & 20 & 13 & 18\\ 
$10\%$ & 1.0 & 0.07 & 0.8 & $75.55\pm1.95$ & $15.18\pm0.04$ & $25.39\pm0.01$ & 2.04 & 0.23 & 17 & 13 & 18\\ 
$10\%$ & 1.0 & 0.10 & 0.5 & $73.10\pm2.05$ & $15.24\pm0.05$ & $25.38\pm0.01$ & 1.68 & 0.21 & 15 & 13 & 18\\ 
$10\%$ & 1.0 & 0.10 & 0.6 & $73.02\pm2.05$ & $15.24\pm0.05$ & $25.38\pm0.01$ & 1.79 & 0.21 & 15 & 13 & 18\\ 
$10\%$ & 1.0 & 0.10 & 0.8 & $72.34\pm2.06$ & $15.27\pm0.05$ & $25.39\pm0.01$ & 2.17 & 0.19 & 14 & 13 & 18\\ 
$10\%$ & 1.0 & 0.15 & 0.5 & $71.85\pm1.99$ & $15.25\pm0.05$ & $25.36\pm0.01$ & 1.70 & 0.19 & 13 & 13 & 18\\ 
$10\%$ & 1.0 & 0.15 & 0.6 & $71.87\pm1.99$ & $15.25\pm0.05$ & $25.36\pm0.01$ & 1.71 & 0.19 & 13 & 13 & 18\\ 
$10\%$ & 1.0 & 0.15 & 0.8 & $71.87\pm1.99$ & $15.25\pm0.05$ & $25.36\pm0.01$ & 1.71 & 0.19 & 13 & 13 & 18\\ 
$10\%$ & 1.5 & 0.07 & 0.5 & $75.89\pm1.90$ & $15.12\pm0.04$ & $25.35\pm0.02$ & 20.85 & 0.23 & 32 & 13 & 18\\ 
$10\%$ & 1.5 & 0.07 & 0.6 & $73.71\pm1.95$ & $15.18\pm0.04$ & $25.35\pm0.02$ & 9.22 & 0.23 & 28 & 13 & 18\\ 
$10\%$ & 1.5 & 0.07 & 0.8 & $72.96\pm1.99$ & $15.21\pm0.05$ & $25.35\pm0.01$ & 1.83 & 0.20 & 20 & 12 & 17\\ 
$10\%$ & 1.5 & 0.10 & 0.5 & $73.90\pm2.15$ & $15.21\pm0.05$ & $25.38\pm0.01$ & 1.55 & 0.23 & 17 & 12 & 17\\ 
$10\%$ & 1.5 & 0.10 & 0.6 & $73.58\pm2.17$ & $15.22\pm0.05$ & $25.38\pm0.01$ & 1.55 & 0.24 & 16 & 12 & 17\\ 
$10\%$ & 1.5 & 0.10 & 0.8 & $71.77\pm2.14$ & $15.24\pm0.05$ & $25.35\pm0.01$ & 1.87 & 0.21 & 14 & 12 & 17\\ 
$10\%$ & 1.5 & 0.15 & 0.5 & $72.77\pm2.10$ & $15.19\pm0.05$ & $25.33\pm0.01$ & 1.53 & 0.25 & 14 & 12 & 17\\ 
$10\%$ & 1.5 & 0.15 & 0.6 & $72.77\pm2.10$ & $15.19\pm0.05$ & $25.33\pm0.01$ & 1.53 & 0.25 & 14 & 12 & 17\\ 
$10\%$ & 1.5 & 0.15 & 0.8 & $72.77\pm2.10$ & $15.19\pm0.05$ & $25.33\pm0.01$ & 1.53 & 0.25 & 14 & 12 & 17\\ 
$10\%$ & 2.0 & 0.07 & 0.5 & $80.33\pm2.25$ & $15.03\pm0.05$ & $25.38\pm0.02$ & 5.95 & 0.25 & 31 & 11 & 16\\ 
$10\%$ & 2.0 & 0.07 & 0.6 & $80.93\pm2.27$ & $15.03\pm0.05$ & $25.39\pm0.01$ & 3.46 & 0.25 & 28 & 11 & 16\\ 
$10\%$ & 2.0 & 0.07 & 0.8 & $74.30\pm2.17$ & $15.19\pm0.05$ & $25.37\pm0.01$ & 1.03 & 0.25 & 22 & 12 & 17\\ 
$10\%$ & 2.0 & 0.10 & 0.5 & $81.00\pm2.42$ & $15.02\pm0.05$ & $25.39\pm0.01$ & 1.55 & 0.25 & 19 & 11 & 16\\ 
$10\%$ & 2.0 & 0.10 & 0.6 & $81.00\pm2.42$ & $15.02\pm0.05$ & $25.39\pm0.01$ & 1.55 & 0.25 & 19 & 11 & 16\\ 
$10\%$ & 2.0 & 0.10 & 0.8 & $73.78\pm2.29$ & $15.20\pm0.06$ & $25.37\pm0.01$ & 1.54 & 0.26 & 16 & 12 & 17\\ 
$10\%$ & 2.0 & 0.15 & 0.5 & $73.26\pm2.25$ & $15.19\pm0.06$ & $25.34\pm0.02$ & 1.10 & 0.27 & 15 & 12 & 17\\ 
$10\%$ & 2.0 & 0.15 & 0.6 & $73.46\pm2.25$ & $15.19\pm0.06$ & $25.34\pm0.02$ & 1.06 & 0.27 & 15 & 12 & 17\\ 
$10\%$ & 2.0 & 0.15 & 0.8 & $73.46\pm2.25$ & $15.19\pm0.06$ & $25.34\pm0.02$ & 1.06 & 0.27 & 15 & 12 & 17\\ 
$20\%$ & 0.75 & 0.07 & 0.5 & $71.15\pm1.91$ & $15.18\pm0.04$ & $25.26\pm0.02$ & 24.04 & 0.20 & 25 & 13 & 18\\ 
$20\%$ & 0.75 & 0.07 & 0.6 & $73.38\pm1.94$ & $15.21\pm0.04$ & $25.36\pm0.01$ & 8.30 & 0.20 & 21 & 13 & 18\\ 
$20\%$ & 0.75 & 0.07 & 0.8 & $74.11\pm1.97$ & $15.22\pm0.05$ & $25.39\pm0.01$ & 2.48 & 0.20 & 19 & 13 & 18\\ 
$20\%$ & 0.75 & 0.10 & 0.5 & $72.67\pm2.03$ & $15.24\pm0.05$ & $25.37\pm0.01$ & 1.17 & 0.20 & 16 & 13 & 18\\ 
$20\%$ & 0.75 & 0.10 & 0.6 & $72.15\pm2.04$ & $15.25\pm0.05$ & $25.37\pm0.01$ & 1.16 & 0.20 & 15 & 13 & 18\\ 
$20\%$ & 0.75 & 0.10 & 0.8 & $72.41\pm2.05$ & $15.25\pm0.05$ & $25.38\pm0.01$ & 1.43 & 0.20 & 15 & 13 & 18\\ 
$20\%$ & 0.75 & 0.15 & 0.5 & $72.92\pm1.98$ & $15.24\pm0.05$ & $25.38\pm0.01$ & 1.81 & 0.19 & 14 & 13 & 18\\ 
$20\%$ & 0.75 & 0.15 & 0.6 & $72.92\pm1.98$ & $15.24\pm0.05$ & $25.38\pm0.01$ & 1.81 & 0.19 & 14 & 13 & 18\\ 
$20\%$ & 0.75 & 0.15 & 0.8 & $72.92\pm1.98$ & $15.24\pm0.05$ & $25.38\pm0.01$ & 1.81 & 0.19 & 14 & 13 & 18\\ 
$20\%$ & 1.0 & 0.07 & 0.5 & $79.79\pm1.93$ & $14.97\pm0.04$ & $25.31\pm0.01$ & 16.59 & 0.23 & 32 & 12 & 17\\ 
$20\%$ & 1.0 & 0.07 & 0.6 & $79.79\pm1.97$ & $15.02\pm0.04$ & $25.35\pm0.01$ & 10.82 & 0.23 & 27 & 12 & 17\\ 
$20\%$ & 1.0 & 0.07 & 0.8 & $76.28\pm1.97$ & $15.15\pm0.04$ & $25.39\pm0.01$ & 1.34 & 0.22 & 19 & 13 & 18\\ 
$20\%$ & 1.0 & 0.10 & 0.5 & $82.84\pm2.31$ & $14.97\pm0.05$ & $25.39\pm0.01$ & 1.17 & 0.25 & 19 & 11 & 14\\ 
$20\%$ & 1.0 & 0.10 & 0.6 & $80.09\pm2.22$ & $15.04\pm0.05$ & $25.38\pm0.01$ & 1.23 & 0.25 & 19 & 12 & 17\\ 
$20\%$ & 1.0 & 0.10 & 0.8 & $74.01\pm2.10$ & $15.22\pm0.05$ & $25.39\pm0.01$ & 1.35 & 0.23 & 17 & 13 & 18\\ 
$20\%$ & 1.0 & 0.15 & 0.5 & $73.24\pm2.05$ & $15.23\pm0.05$ & $25.37\pm0.01$ & 2.14 & 0.21 & 14 & 13 & 18\\ 
$20\%$ & 1.0 & 0.15 & 0.6 & $73.24\pm2.05$ & $15.23\pm0.05$ & $25.37\pm0.01$ & 2.14 & 0.21 & 14 & 13 & 18\\ 
$20\%$ & 1.0 & 0.15 & 0.8 & $73.24\pm2.05$ & $15.23\pm0.05$ & $25.37\pm0.01$ & 2.14 & 0.21 & 14 & 13 & 18\\ 
$20\%$ & 1.5 & 0.07 & 0.5 & $81.10\pm2$ & $14.96\pm0.04$ & $25.33\pm0.02$ & 9.07 & 0.24 & 34 & 12 & 17\\ 
$20\%$ & 1.5 & 0.07 & 0.6 & $73.90\pm1.99$ & $15.18\pm0.04$ & $25.35\pm0.02$ & 4.11 & 0.21 & 29 & 13 & 18\\ 
$20\%$ & 1.5 & 0.07 & 0.8 & $73.26\pm2.10$ & $15.19\pm0.05$ & $25.34\pm0.01$ & 0.97 & 0.22 & 19 & 12 & 17\\ 
$20\%$ & 1.5 & 0.10 & 0.5 & $74.59\pm2.18$ & $15.17\pm0.05$ & $25.36\pm0.01$ & 1.20 & 0.26 & 18 & 12 & 17\\ 
$20\%$ & 1.5 & 0.10 & 0.6 & $74.67\pm2.18$ & $15.17\pm0.05$ & $25.36\pm0.01$ & 1.26 & 0.26 & 18 & 12 & 17\\ 
$20\%$ & 1.5 & 0.10 & 0.8 & $72.54\pm2.16$ & $15.20\pm0.05$ & $25.33\pm0.01$ & 1.70 & 0.23 & 16 & 12 & 17\\ 
$20\%$ & 1.5 & 0.15 & 0.5 & $73\pm2.13$ & $15.17\pm0.05$ & $25.32\pm0.02$ & 1.35 & 0.25 & 15 & 12 & 17\\ 
$20\%$ & 1.5 & 0.15 & 0.6 & $73\pm2.13$ & $15.17\pm0.05$ & $25.32\pm0.02$ & 1.35 & 0.25 & 15 & 12 & 17\\ 
$20\%$ & 1.5 & 0.15 & 0.8 & $73\pm2.13$ & $15.17\pm0.05$ & $25.32\pm0.02$ & 1.35 & 0.25 & 15 & 12 & 17\\ 
$20\%$ & 2.0 & 0.07 & 0.5 & $81.05\pm2.15$ & $14.98\pm0.04$ & $25.35\pm0.01$ & 4.62 & 0.24 & 35 & 12 & 17\\ 
$20\%$ & 2.0 & 0.07 & 0.6 & $78.24\pm2.01$ & $15.07\pm0.04$ & $25.36\pm0.01$ & 3.48 & 0.27 & 32 & 13 & 18\\ 
$20\%$ & 2.0 & 0.07 & 0.8 & $76.12\pm2.20$ & $15.14\pm0.05$ & $25.37\pm0.01$ & 1.02 & 0.30 & 25 & 13 & 18\\ 
$20\%$ & 2.0 & 0.10 & 0.5 & $86.22\pm2.53$ & $14.88\pm0.05$ & $25.38\pm0.02$ & 1.18 & 0.27 & 22 & 11 & 14\\ 
$20\%$ & 2.0 & 0.10 & 0.6 & $83.13\pm2.42$ & $14.96\pm0.05$ & $25.38\pm0.02$ & 1.17 & 0.28 & 22 & 12 & 17\\ 
$20\%$ & 2.0 & 0.10 & 0.8 & $74.78\pm2.29$ & $15.15\pm0.06$ & $25.34\pm0.02$ & 1.45 & 0.30 & 19 & 13 & 18\\ 
$20\%$ & 2.0 & 0.15 & 0.5 & $73.30\pm2.28$ & $15.20\pm0.06$ & $25.35\pm0.02$ & 1.62 & 0.24 & 15 & 12 & 17\\ 
$20\%$ & 2.0 & 0.15 & 0.6 & $73.03\pm2.27$ & $15.20\pm0.06$ & $25.34\pm0.02$ & 1.77 & 0.24 & 15 & 12 & 17\\ 
$20\%$ & 2.0 & 0.15 & 0.8 & $73.03\pm2.27$ & $15.20\pm0.06$ & $25.34\pm0.02$ & 1.77 & 0.24 & 15 & 12 & 17\\ 
$5\%$ & 0.75 & 0.07 & 0.5 & $70.38\pm1.69$ & $15.22\pm0.04$ & $25.28\pm0.01$ & 13.40 & 0.19 & 23 & 13 & 18\\ 
$5\%$ & 0.75 & 0.07 & 0.6 & $70.78\pm1.72$ & $15.23\pm0.04$ & $25.31\pm0.01$ & 15.54 & 0.19 & 21 & 13 & 18\\ 
$5\%$ & 0.75 & 0.07 & 0.8 & $72.70\pm1.80$ & $15.25\pm0.04$ & $25.38\pm0.01$ & 0.70 & 0.20 & 16 & 13 & 18\\ 
$5\%$ & 0.75 & 0.10 & 0.5 & $72.72\pm1.84$ & $15.25\pm0.04$ & $25.39\pm0.01$ & 1.37 & 0.20 & 16 & 13 & 18\\ 
$5\%$ & 0.75 & 0.10 & 0.6 & $72.77\pm1.84$ & $15.25\pm0.04$ & $25.39\pm0.01$ & 1.49 & 0.20 & 16 & 13 & 18\\ 
$5\%$ & 0.75 & 0.10 & 0.8 & $73.17\pm1.86$ & $15.26\pm0.04$ & $25.41\pm0.01$ & 1.70 & 0.20 & 15 & 13 & 18\\ 
$5\%$ & 0.75 & 0.15 & 0.5 & $71.67\pm1.84$ & $15.26\pm0.04$ & $25.36\pm0.01$ & 1.32 & 0.18 & 13 & 13 & 18\\ 
$5\%$ & 0.75 & 0.15 & 0.6 & $71.67\pm1.84$ & $15.26\pm0.04$ & $25.36\pm0.01$ & 1.32 & 0.18 & 13 & 13 & 18\\ 
$5\%$ & 0.75 & 0.15 & 0.8 & $71.93\pm1.85$ & $15.26\pm0.04$ & $25.37\pm0.01$ & 1.44 & 0.18 & 13 & 13 & 18\\ 
$5\%$ & 1.0 & 0.07 & 0.5 & $72.61\pm1.81$ & $15.22\pm0.04$ & $25.35\pm0.01$ & 10.43 & 0.18 & 21 & 13 & 18\\ 
$5\%$ & 1.0 & 0.07 & 0.6 & $73.35\pm1.85$ & $15.24\pm0.04$ & $25.40\pm0.01$ & 4.95 & 0.19 & 20 & 13 & 18\\ 
$5\%$ & 1.0 & 0.07 & 0.8 & $73.46\pm1.91$ & $15.26\pm0.04$ & $25.41\pm0.01$ & 2.05 & 0.21 & 15 & 13 & 18\\ 
$5\%$ & 1.0 & 0.10 & 0.5 & $71.64\pm1.95$ & $15.28\pm0.05$ & $25.38\pm0.01$ & 5.48 & 0.19 & 14 & 13 & 18\\ 
$5\%$ & 1.0 & 0.10 & 0.6 & $71.74\pm1.95$ & $15.28\pm0.05$ & $25.38\pm0.01$ & 5.36 & 0.19 & 14 & 13 & 18\\ 
$5\%$ & 1.0 & 0.10 & 0.8 & $71.72\pm1.96$ & $15.28\pm0.05$ & $25.38\pm0.01$ & 6.34 & 0.19 & 13 & 13 & 18\\ 
$5\%$ & 1.0 & 0.15 & 0.5 & $71.53\pm1.93$ & $15.26\pm0.05$ & $25.35\pm0.01$ & 1.59 & 0.19 & 13 & 13 & 18\\ 
$5\%$ & 1.0 & 0.15 & 0.6 & $71.53\pm1.93$ & $15.26\pm0.05$ & $25.35\pm0.01$ & 1.59 & 0.19 & 13 & 13 & 18\\ 
$5\%$ & 1.0 & 0.15 & 0.8 & $71.53\pm1.93$ & $15.26\pm0.05$ & $25.35\pm0.01$ & 1.59 & 0.19 & 13 & 13 & 18\\ 
$5\%$ & 1.5 & 0.07 & 0.5 & $73.07\pm1.87$ & $15.22\pm0.04$ & $25.36\pm0.01$ & 5.74 & 0.19 & 27 & 13 & 18\\ 
$5\%$ & 1.5 & 0.07 & 0.6 & $73.08\pm1.88$ & $15.23\pm0.04$ & $25.37\pm0.01$ & 2.70 & 0.19 & 25 & 13 & 18\\ 
$5\%$ & 1.5 & 0.07 & 0.8 & $67.91\pm2.08$ & $15.36\pm0.06$ & $25.35\pm0.01$ & 1.35 & 0.09 & 15 & 11 & 16\\ 
$5\%$ & 1.5 & 0.10 & 0.5 & $73.68\pm2.10$ & $15.24\pm0.05$ & $25.40\pm0.01$ & 1.42 & 0.21 & 15 & 12 & 17\\ 
$5\%$ & 1.5 & 0.10 & 0.6 & $72.05\pm2.12$ & $15.27\pm0.05$ & $25.39\pm0.01$ & 1.48 & 0.20 & 13 & 12 & 17\\ 
$5\%$ & 1.5 & 0.10 & 0.8 & $71.06\pm2.09$ & $15.26\pm0.05$ & $25.34\pm0.01$ & 1.71 & 0.20 & 12 & 12 & 17\\ 
$5\%$ & 1.5 & 0.15 & 0.5 & $70.53\pm2.03$ & $15.24\pm0.05$ & $25.31\pm0.02$ & 1.41 & 0.20 & 12 & 12 & 17\\ 
$5\%$ & 1.5 & 0.15 & 0.6 & $70.53\pm2.03$ & $15.24\pm0.05$ & $25.31\pm0.02$ & 1.41 & 0.20 & 12 & 12 & 17\\ 
$5\%$ & 1.5 & 0.15 & 0.8 & $70.53\pm2.03$ & $15.24\pm0.05$ & $25.31\pm0.02$ & 1.41 & 0.20 & 12 & 12 & 17\\ 
$5\%$ & 2.0 & 0.07 & 0.5 & $75.85\pm1.97$ & $15.16\pm0.04$ & $25.38\pm0.01$ & 5.25 & 0.24 & 29 & 13 & 18\\ 
$5\%$ & 2.0 & 0.07 & 0.6 & $75.89\pm2$ & $15.17\pm0.04$ & $25.40\pm0.02$ & 3.63 & 0.23 & 25 & 13 & 18\\ 
$5\%$ & 2.0 & 0.07 & 0.8 & $71.64\pm2.11$ & $15.26\pm0.05$ & $25.36\pm0.01$ & 6.84 & 0.21 & 15 & 12 & 17\\ 
$5\%$ & 2.0 & 0.10 & 0.5 & $72.83\pm2.23$ & $15.24\pm0.06$ & $25.37\pm0.02$ & 0.60 & 0.21 & 17 & 12 & 17\\ 
$5\%$ & 2.0 & 0.10 & 0.6 & $71.92\pm2.26$ & $15.26\pm0.06$ & $25.37\pm0.02$ & 0.60 & 0.22 & 15 & 12 & 17\\ 
$5\%$ & 2.0 & 0.10 & 0.8 & $70.52\pm2.22$ & $15.26\pm0.06$ & $25.32\pm0.02$ & 0.60 & 0.21 & 13 & 12 & 17\\ 
$5\%$ & 2.0 & 0.15 & 0.5 & $70.25\pm2.17$ & $15.25\pm0.06$ & $25.31\pm0.02$ & 1.10 & 0.21 & 12 & 12 & 17\\ 
$5\%$ & 2.0 & 0.15 & 0.6 & $70.19\pm2.17$ & $15.25\pm0.06$ & $25.31\pm0.02$ & 1.11 & 0.21 & 12 & 12 & 17\\ 
$5\%$ & 2.0 & 0.15 & 0.8 & $70.19\pm2.17$ & $15.25\pm0.06$ & $25.31\pm0.02$ & 1.11 & 0.21 & 12 & 12 & 17\\
\enddata
\end{deluxetable*}

\null\newpage
\null\newpage

\section{Four SN Ia Hosts which lack a significant tip detection}

There are four galaxies, NGC 5584, 3021, 3370, and 1309, which are also the most distant of the sample, with expected $32 < \mu < 32.7$ mag based on their SNe Ia (and Cepheids), for which we do not detect a significant tip.  The HST observations of these galaxies targeted the disk in order to measure Cepheid variables \cite{Riess22}, making it doubly difficult to measure their TRGBs.  TRGB measurements via edge detection have been obtained by \cite{Jang_2018}, with tip magnitudes $m_I \sim 28$ and formal uncertainties of $\leq$ 0.1 mag.  These values were retained in the analyses of \cite{Freedman19b} and \cite{Freedman21}.    \cite{Anand22} remeasured the photometry for these hosts and determined that the CMDs were not deep enough to measure their TRGBs, with the expected location of the tip approaching SNR $\sim$ 3, in good agreement with the HST exposure time calculator, so this conclusion is not sensitive to the method used to measure the photometry.

\begin{figure*}
    \centering
   
\includegraphics[width=1.0\textwidth]{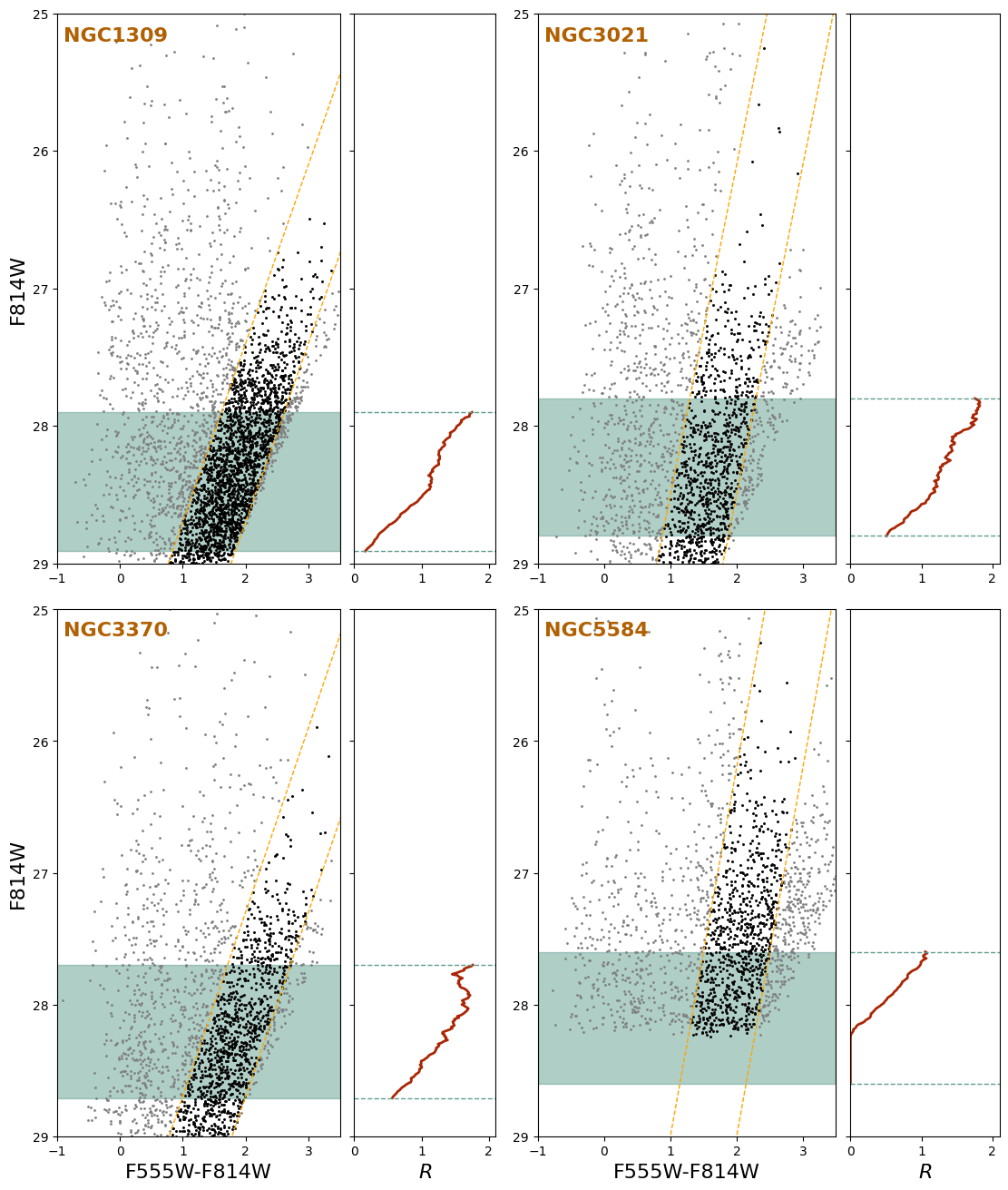}
    \caption{The CMDs and Contrast Ratio versus F814W magnitude plot for NGC1309, NGC3021, NGC3370 and NGC5584. The shaded green region is the expected TRGB in the F814W filter as shown by the curve on the right of each CMD, the value of the contrast ratio $ R $ at those magnitudes never exceeds 2. } 
    \label{fig:EDResponse}
\end{figure*}

We show the CMDs for these galaxies in Fig.~\ref{fig:EDResponse}.  We fail to detect a significant tip in a broad, one magnitude range centered on the expected value. A more rigorous metric, the value of $R$, never exceeds 2 across that interval (right side of each panel). Therefore, even if a tip was found in this range, the analysis in W22 has empirically demonstrated that the uncertainty of tips with such little contrast ($R<2$) is $>$ 1 magnitude, making such a measurement meaningless.  We therefore follow the decision of \cite{Anand22} to remove these galaxies from our sample.  It is incumbent upon any future analysis that seeks to make use of TRGB measurements in these hosts to provide the data and tools necessary to demonstrate that these hosts yield a significant TRGB measurement.

\end{appendix}

\end{document}